\documentclass[traditabstract]{aa}
\usepackage{xspace}
\usepackage[mathscr]{eucal}
\usepackage{array}
\usepackage{morefloats}
\usepackage{multirow}
\usepackage{enumerate}
\usepackage{natbib}
\usepackage[stable]{footmisc}
\usepackage{amsmath}
\usepackage{amssymb}
\usepackage[varg]{txfonts}
\usepackage{placeins}
\usepackage{natbib}
\usepackage[utf8]{inputenc}

\bibpunct{(}{)}{;}{a}{}{,} %
\usepackage{graphicx}
\usepackage{epstopdf}
\usepackage{ifthen}
\usepackage[table,usenames,dvipsnames]{xcolor}
\definecolor{linkcolor}{rgb}{0.6,0,0}
\definecolor{citecolor}{rgb}{0,0,0.75}
\definecolor{urlcolor}{rgb}{0.12,0.46,0.7}
\usepackage[breaklinks, colorlinks, urlcolor=urlcolor, linkcolor=linkcolor,citecolor=citecolor,pdfencoding=auto]{hyperref}
\usepackage{overpic}
\usepackage[normalem]{ulem}
\usepackage{rotating}
\usepackage{booktabs}
\usepackage{subcaption}
\usepackage{ulem}
\usepackage{etoolbox}
\usepackage{listings}
\usepackage{lipsum}
\usepackage[low-sup]{subdepth}
\usepackage{tikz}
\usepackage{cuted}
\usetikzlibrary{positioning}
\usetikzlibrary{shapes.geometric}
\usetikzlibrary{backgrounds}

\usepackage{ulem}

\newcommand{\planck}{\textit{Planck}\xspace}

\providecommand{\cov}{\text{cov}}
\providecommand{\half}{\frac{1}{2}}

 \newcommand{\hC}{\ensuremath{\hat{C}}}
\newcommand{\tC}{\ensuremath{\tilde{C}}}
\newcommand{\ta}{\ensuremath{\tilde{a}}}
\newcommand{\tSig}{\ensuremath{\tilde{\Sigma}}}
\newcommand{\hSig}{\ensuremath{\hat{\Sigma}}}
\newcommand{\bThe}{\ensuremath{\bar{\Theta}}}
\newcommand{\hu}{\ensuremath{}\hat{u}} \newcommand{\hn}{\ensuremath{}\hat{n}}
\newcommand{\nside}{\ensuremath{n_{\rm side}}}
\newcommand{\Nsim}{\ensuremath{N_\mathrm{sim}}} \newcommand{\sims}{\mksym{{\rm
sim}}\xspace} \newcommand{\fapo}{\ensuremath{f^\mathrm{apo}}}
\newcommand{\ie}{\emph{i.e.}\xspace} \newcommand{\lhs}{left hand side\xspace}
\newcommand{\rhs}{right hand side\xspace}

\newcommand{\act}{{ACT}\xspace} \newcommand{\spt}{\textsc{SPT-3G}\xspace}
\newcommand{\sptyeete}{\textsc{SPT-3G}\xspace}
    \providecommand{\cmb}{\textsc{CMB}\xspace}

\providecommand{\healpix}{\texttt{HEALPix}\xspace}

\providecommand{\maptoalm}{\texttt{map2alm}\xspace}

\providecommand{\almtomap}{\texttt{alm2map}\xspace}
\providecommand{\polspice}{\texttt{PolSpice}\xspace}

\providecommand{\master}{\texttt{MASTER}\xspace}
\providecommand{\mx}{\mathrm{max}}

\newcommand{\lmax}{\ensuremath{\ell_{\mathrm{max}}}}

\newcommand{\lexact}{\ensuremath{\ell_{\rm max, ex}}}
\newcommand{\lcut}{\ensuremath{\ell_{\mathrm{cut}}}}

\newcommand{\mksym}[1]{\ifmmode {\rm #1}\else #1\fi}
 \newcommand{\TT}{\mksym{{\rm
\textsc{TT}}}\xspace} \newcommand{\EE}{\mksym{{\rm \textsc{EE}}}\xspace}
\newcommand{\BB}{\mksym{{\rm \textsc{BB}}}\xspace} \newcommand{\TE}{\mksym{{\rm
\textsc{TE}}}\xspace} 
  \newcommand{\T}{\mksym{{\rm \textsc{T}}}\xspace}
  \newcommand{\E}{\mksym{{\rm \textsc{E}}}\xspace}
\newcommand{\B}{\mksym{{\rm \textsc{B}}}\xspace}

\newcommand{\TTTT}{\textsc{TTTT}\xspace}
\newcommand{\EEBB}{\textsc{EEBB}\xspace}
\newcommand{\BBEE}{\textsc{BBEE}\xspace}
\newcommand{\BBBB}{\textsc{BBBB}\xspace}
\newcommand{\EEEE}{\textsc{EEEE}\xspace}

\newcommand{\INKA}{\textsc{INKA}\xspace} \newcommand{\FRI}{\textsc{FRI}\xspace}
\newcommand{\NKA}{\textsc{NKA}\xspace} \newcommand{\ACC}{\textsc{ACC}\xspace}

\newcommand{\begm}{\begin{pmatrix}} \newcommand{\enm}{\end{pmatrix}}

\newcommand{\ud}{{\rm d}}

\def\beglet{ \addtocounter{equation}{1}%
  \setcounter{parentequation}{\value{equation}}%
  \setcounter{equation}{0}%
  \def\theequation{\arabic{parentequation}\alph{equation}}%
  \ignorespaces
} \def\endlet{ \setcounter{equation}{\value{parentequation}}%
  \def\theequation{\arabic{equation}}%
}

\providecommand{\beglet}{\begin{subequations}}
    \providecommand{\endlet}{\end{subequations}}

\makeatother

\begin{document}
% \linenumbers

\title{Accurate CMB covariance matrices: exact calculation and approximations}

\author{E. Camphuis\inst{\ref{inst1}} \and  K. Benabed\inst{\ref{inst1}} \and S.
  Galli\inst{\ref{inst1}} \and E. Hivon\inst{\ref{inst1}} \and M.
  Lilley\inst{\ref{inst2}}}

\institute{ Sorbonne Universit\'{e}, UMR7095, Institut d'Astrophysique de Paris,
98 bis Boulevard Arago, F-75014, Paris, France\label{inst1} \and SYRTE,
Observatoire de Paris, Université PSL, CNRS, Sorbonne Université, LNE, 61 avenue
de l’Observatoire 75014 Paris, France\label{inst2} }

\abstract{\vglue -3mm Accurate covariance matrices are required for a reliable
    estimation of cosmological parameters from pseudo-power spectrum estimators.
    In this work, we focus on the analytical calculation of covariance matrices.
    We consider the case of observations of the Cosmic Microwave Background in
    temperature and polarization on a small footprint such as in the \spt
    experiment, which observes $4\%$ of the sky. Power spectra evaluated on
    small footprints are expected to have large correlations between modes, and
    these need to be accurately modelled. We present, for the first time, an
    algorithm that allows an efficient (but computationally expensive) exact
    calculation of analytic covariance matrices. Using it as our reference, we
    test the accuracy of existing fast approximations of the covariance matrix.
    We find that, when the power spectrum is binned in wide bandpowers, current
    approaches are correct up to the 5\% level on the \spt small sky footprint.
    Furthermore, we propose a new approximation which improves over the previous
    ones reaching a precision of 1\% in the wide bandpowers case and generally
    more than 4 times more accurate than current approaches. Finally, we derive
    the covariance matrices for mask-corrected power spectra estimated by the
    \polspice code. In particular, we include, in the case of a small sky
    fraction, the effect of the apodization of the large scale modes. While we
    considered the specific case of the CMB, our results are applicable to any
    other cosmological probe which requires the calculation of pseudo-power
    spectrum covariance matrices.}


\authorrunning{E. Camphuis et al.}
\titlerunning{Accurate CMB covariance matrices}

\maketitle

% \tableofcontents

%%%%%%%%%%%%%%%%%%%%%%
\section{Introduction}
One of the most powerful probes of cosmology is the observation of the Cosmic
Microwave Background (\cmb) anisotropies. The \textsc{ESA} Planck satellite \cmb
measurements marked the entry into the era of precision cosmology, with many
$\Lambda$CDM cosmological parameters measured with uncertainties smaller than
$1\%$ \citep{Aghanim:2018eyx}. Ongoing and upcoming ground-based and satellite
experiments such as the Atacama Cosmology Telescope (\act)
\citep{Aiola:2020azj}, the South Pole Telescope (\textsc{SPT})
\citep{Dutcher2021}, Simons Observatory (\textsc{SO})
\citep{2019JCAP...02..056A}, \cmb-Stage 4 (\cmb-\textsc{S4})
\citep{cmbs4-sciencebook} and Litebird \citep{hazumi2012} will provide yet more
information about the nature of our universe.

Since primary \cmb anisotropies in intensity and polarization are distributed as
a gaussian random field, most of the cosmological information is contained in
the angular power spectrum of the \cmb anisotropies. As the evolution of the
primary anisotropies is linear, the multipoles of the angular power spectrum are
uncorrelated when observing the full sky. However, any realistic experiment
requires masking parts of the sky, either to avoid regions highly contaminated
by foregrounds (such as galactic emission or point sources), or because the
scanning strategy is designed to observe specific regions of the sky. The
estimation of the power spectrum on the masked sky, the so-called pseudo-power
spectrum, is biased and different multipoles become correlated \citep{Hietal02}.
An unbiased estimator of the spectra can then be obtained through the \master
approach \citep{Hietal02}, as implemented e.g. in the \polspice
\footnote{\url{http://www2.iap.fr/users/hivon/software/PolSpice/}} software
\citep{Szapudi2001, Chonetal04}. A robust inference of cosmological parameters
requires accurate covariance matrices that describe the variance of the spectra
along their diagonal, as well as the correlations between multipoles in the
off-diagonal terms. Pseudo-$C_\ell$ covariance matrices are corrected for the
effect of the mask using \master to obtain the covariance matrices for the
unbiased $C_\ell$ estimator. Inaccuracies in the covariance matrix estimation
can lead to the misestimation of cosmological parameters and of their
uncertainties \citep{Dodelson2013,Sellentin2019}.

Covariance matrices can be calculated through the use of simulations. The number
of simulations determines the accuracy of the estimator. As the simulations are
expensive to produce, the obtained noisy realization of the covariance has to be
regularized \citep{balkenhol2021}\footnote{While this work focuses on covariance
estimates obtained through empirical estimators, the conditioning schemes it
presents can similarly be applied to estimates from simulations.}.
Alternatively, it is possible to calculate pseudo-$C_\ell$ covariance matrices
analytically. However, these depend on integrals whose exact numerical
implementation is computationally expensive. Thus, approximations have been
proposed in previous works to make these calculations efficient, see e.g.
\citet{Efstathiou:2004,Nicola2021,Friedrich2021}.

We analyze the problem of computing accurate analytical
covariance matrices. We take the specific case of the \spt experiment, which
observes the \cmb anisotropies in temperature and polarization on a small sky
patch, which corresponds to about 4\% of the sky. On such a small sky region,
the calculated power spectra has large correlations between multipoles. The
existing approximations of the covariance matrix can be less accurate in these
conditions. Considering this particular case is thus a particularly stringent
test of the validity of analytical algorithms.

We implement for the first time the exact computationally expensive calculation
of the covariance matrices, which we find to be numerically feasible at
multipoles smaller than $\ell\lesssim \lexact \equiv 1000$ thanks to a new
algorithm that gains one order of numerical complexity over the brute-force
approach, resulting in a thousand-fold speed improvement. Then, we test the
existing approximations, and find that they are accurate at the 5\% level in the
case of the \spt footprint when the power spectrum is averaged in wide
bandpowers. We then propose a new approximation which improves over the existing
algorithms to attain an accuracy of 1\% in the same case. Finally, we describe
how to calculate the covariance matrix of the \polspice $C_\ell$ estimator from
the pseudo-$C_\ell$ covariance matrix.

While in this work we focused on the specific case of the \spt \cmb experiment,
many considerations can be applied more broadly to any probe relying on the
calculation of power spectra and covariance matrices.

The paper is organized as follows. In Sect.~\ref{sec:pseudocov}, we introduce
the pseudo-power spectrum estimator and its covariance. In Sect.~\ref{sec:exact}
we show how to perform the exact calculation of the covariance matrix. In
Sect.~\ref{sec:oldapprox} we describe the existing approximations for the
calculation of the covariance matrix  and we test their accuracy against the
exact computation. Section~\ref{sec:newapprox} presents our new approximation
which is more accurate. Sect.~\ref{sec:polspice} describes how to calculate the
covariance matrix of the \polspice estimator. We conclude in
Sect.~\ref{sec:conclusions}, and some detailed calculations are given in the
appendices.
%%%%%%%%%%%%%%%%%%%%%%

%%%%%%%%%%%%%%%%%%%%%%
\section{Covariance of the pseudo-power spectrum}
\label{sec:pseudocov}
\subsection{Pseudo-power spectrum}
\cmb anisotropies in intensity and polarization can be described as maps of
Stokes parameters $T(\hat{n}), Q(\hat{n}), U(\hat{n})$ for a direction $\hat{n}$
of the sky. They are gaussian random fields, fully characterized by their
angular power spectra $(C_\ell^\TT, C_\ell^\EE, C_\ell^\BB, C_\ell^\TE)$, which
are the variances of the harmonic coefficients $a^\T_{\ell m}, a^\E_{\ell m},
a^\B_{\ell m}$ obtained by spherical harmonic decomposition of the maps.
Cosmological models allow the computation of the expectation of the different
power spectra in an ideal full-sky case. However, data only ever covers a part
of the sky. We describe the partial coverage with the weight map $W(\hat{n})$.
The power spectrum of masked maps, labelled $\tC_\ell^{\rm XY}$, is usually
called \emph{pseudo-power spectrum}. Its expression for temperature is given in
Eq.~(\ref{eq:definepseudoclTT}). It can be computed from the masked harmonic
coefficients $\ta^{\rm X}_{\ell m}$ which are directly related to the unmasked
ones, $a^{\rm X}_{\ell'm'}$, by the application of the mode coupling kernels
${}_{s}I_{\ell m\ell'm'}[W]$. In the case of temperature, we write
\begin{equation}
    \label{eq:tildea_I_alm}
    \ta^\T_{\ell m} = \sum_{\ell'm'} a^\T_{\ell'm'} \
    {}_{0}I_{\ell m\ell'm'}[W],
\end{equation}
where we have defined the mode coupling kernels\footnote{The complex conjugate
    is denoted with a star.}
\begin{equation}
    \label{eq:polIkernels}
    {}_{s}I_{\ell m \ell' m'}[W] \equiv \int
    \ud \hu {}_{s}Y_{\ell m}(\hu)
    W(\hu) {}_{s}Y^*_{\ell' m'}(\hu).
\end{equation}
These coupling kernels are an important component of the following
discussions\footnote{This coupling matrix is often denoted $K$ in the literature
such as in \cite{Hietal02}. In this work, we modified the notation for it to be
consistent with the notation of Sect.~\ref{sec:polspice}.}. In the full-sky
case, the closing relations of spin-weighted spherical harmonics will ensure
that ${}_{s}I_{\ell m\ell'm'}[1]=\delta_{\ell \ell'}\delta_{mm'}$. We recall in
App.~\ref{app:Iprop} some summation properties of products of coupling matrices
that appear in the computation of pseudo-power spectra and their covariance. In
particular, they are related to the symmetric coupling kernel acting on a power
spectrum $\mathcal{A}$, labelled $\Xi[\mathcal{A}]$, with
\begin{equation}
    \Xi^{ss'}_{\ell\ell'}[\mathcal{A}] \equiv
    \sum_L \frac{2L+1}{4\pi}  \mathcal{A}_L
    \begin{pmatrix}
        \ell & \ell' & L \\
        s    & -s    & 0
    \end{pmatrix}
    \begin{pmatrix}
        \ell & \ell' & L \\
        s'   & -s'   & 0
    \end{pmatrix}.
\end{equation}
This operator, introduced in \cite{Efstathiou:2004}, can also be seen as acting
on a map $A$ with power spectrum $\mathcal{A}_\ell$. In the following, we will
use the notation $\Xi[A] \equiv \Xi[\mathcal{A}]$. We recall that the average of
the pseudo-spectrum is related to the underlying power spectrum by the
application of the asymmetric coupling kernel computed for the mask $W$, also
known as the \master mode-coupling matrix $M$. In the case of temperature, we
have
\begin{align}
    \langle \tC^\TT_\ell\rangle & = \sum_{\ell'} {}_0M_{\ell\ell'}
    [W]C^\TT_{\ell'}                                                         \\
    {}_0 M_{\ell\ell'}[W]       & \equiv (2\ell'+1)\Xi^{00}_{\ell \ell'}[W].
\end{align}
In this work, without loss of generality, we will develop the computations for
the intensity case (\ie $s=s'=0$), the polarization case being similar. We will
also assume that a single mask is used for both temperature and polarization.
When required, we will highlight the differences between the temperature and
polarization cases and give insight on the importance of the single mask
assumption.

\subsection{Covariance}
Estimating the covariance of the measured power-spectrum is crucial to assess
the agreement between data and model predictions and to constrain cosmological
parameters from CMB maps. As discussed in \cite{Hietal02} and demonstrated in
App.~\ref{sec:analysis}, masking breaks statistical isotropy and induces
correlations between the modes of the pseudo-spectrum. The details of the
derivation of the analytical expression of the pseudo-spectrum covariance can be
found in App.~\ref{app:covar}. We give here the expression in terms of the
coupling matrices ${}_0I$ and the true underlying intensity power spectrum
$C_\ell$, for the temperature case,
\begin{align}
    \tSig_{\ell \ell'} \equiv & \
    \cov({\tilde{C}_\ell, \tilde{C}_{\ell'}}),    \nonumber                                    \\
    \label{eq:pseudocov_fourI}
    =                         & \frac{2}{(2\ell+1)(2\ell'+1)} \sum_{mm'}
    \sum_{\ell_1m_1} \sum_{\ell_2m_2}
    C_{\ell_1}C_{\ell_2}                                                                       \\ \nonumber
                              & {}_{0}I_{\ell m\ell_1m_1}[W] \ {}_{0}I^*_{\ell'm'\ell_1m_1}[W]
    \ {}_{0}I^*_{\ell m\ell_2m_2}[W] \ {}_{0}I_{\ell'm'\ell_2m_2}[W].
\end{align}
As shown in Eq.~(\ref{eq:tildea_I_alm}), the mode-coupling coefficients
${}_{0}I$ kernels relate the underlying harmonic coefficients to the harmonic
coefficients measured on the sky through the mask. In the analytic expression of
the covariance, they represent the coupling between modes due to partial sky
coverage. An expression similar to Eq.~(\ref{eq:pseudocov_fourI}) can be written
for polarization, using spin-2 spherical harmonics, \ie $s, s' = \pm2$. Those
expressions will mix the \EE and \BB power spectra.

The expression in Eq.~(\ref{eq:pseudocov_fourI}) involves several convolutions
and its evaluation is computationally expensive. The full computation scales as
$\mathcal{O}(\lmax^6)$, \lmax being the largest multipole, making the exact
computation of this covariance a daunting task given the currently available
computation power. We have developed an algorithm that allows the computation of
the covariance matrix at low multipoles with a gain of an order of magnitude in
computational time. We discuss this result in Sect.~\ref{sec:exact}.

\begin{figure}
    \centering
    \includegraphics{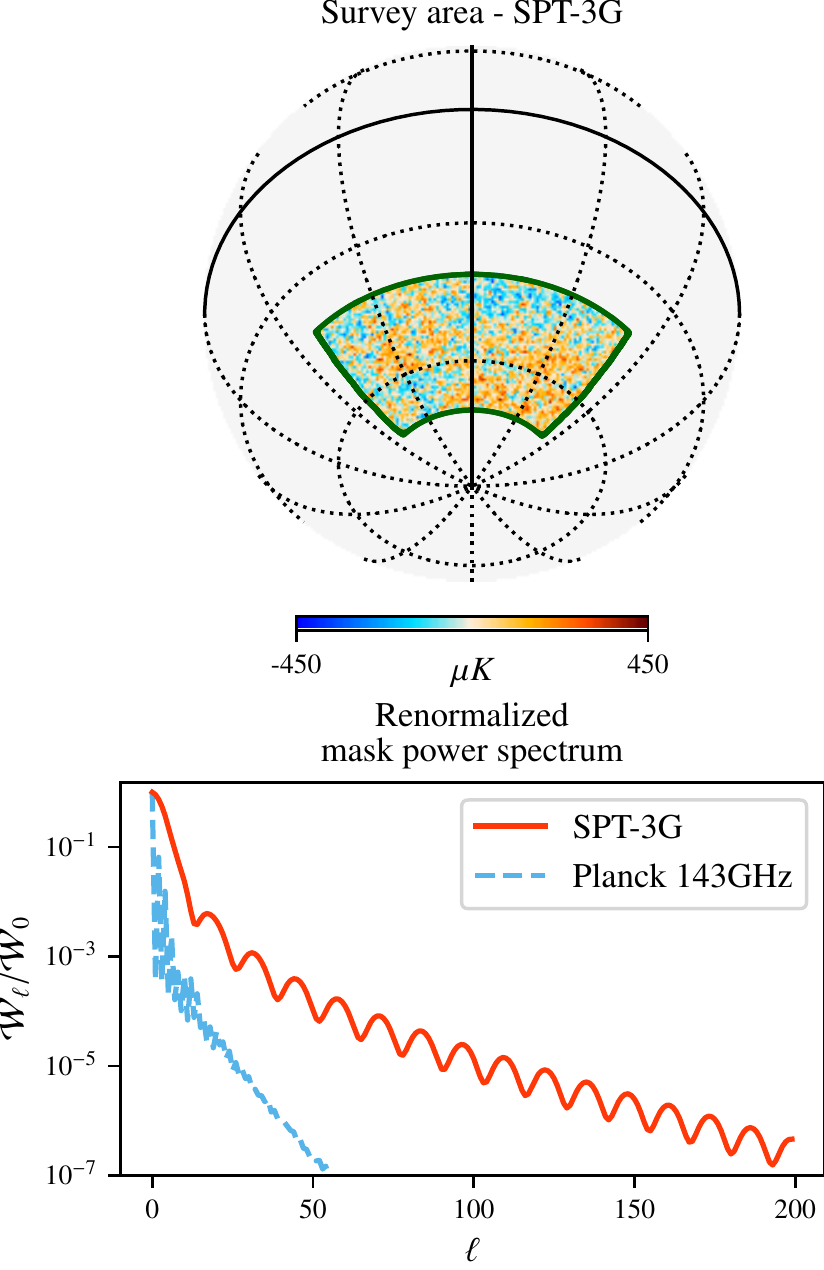}
    \caption{Top: \cmb temperature anisotropies on the \sptyeete patch in
        galactic coordinates. The dark green line delimits the survey footprint.
        The vertical and horizontal bold black lines are the zero-longitude and
        zero-latitude coordinates, respectively. The \sptyeete patch covers
        roughly $4\%$ of the sky. \\  Bottom: Mask power spectra as defined in
        Eq.~(\ref{eq:maskps}) for \sptyeete and for the 143 GHz map used in the
        \planck cosmological analysis, which covers around $70\%$ of the sky.
        The spectra have been renormalised by their first value for comparison
        purposes. Masks corresponding to small sky fractions, such as the \spt
        one, have a shallower power spectrum compared to large ones.}
    \label{fig:maskonsphere}
\end{figure}

With such an approach previously unavailable, existing work has relied on
approximations of Eq.~(\ref{eq:pseudocov_fourI}). In Sect.~\ref{sec:oldapprox}
we will present different approximations that have been proposed in previous
works and we will then validate them against our full computation. This
validation will be performed for a small survey footprint, where spectral modes
are highly correlated. These correlations can challenge the assumptions made in
the different approximations. Throughout this work, we use a test-case inspired
by \spt. The footprint of the first year survey presented in \cite{Dutcher2021}
covers roughly $4\%$ of the sky and is displayed in Fig.~\ref{fig:maskonsphere}
along with the mask power spectrum $\mathcal{W}_\ell$. We apodize the mask with
a Gaussian window function of $30$ arcmin full width half maximum, using an
algorithm similar to the one used in \planck \cite{planck2018likelihood}.We also
show in Fig.~\ref{fig:maskonsphere} the power spectrum of one of the masks used
in the \planck cosmological analysis, which covers a much larger patch of the
sky, around $70\%$ before apodization. The precision of the standard
approximation of the covariance was validated in the latter case, but it needs
to be assessed for a smaller survey area.

%%%%%%%%%%%%%%%%%%%%%%

%%%%%%%%%%%%%%%%%%%%%%
\section{Exact computation}
%%%
\label{sec:exact}
An exact calculation of the pseudo-power spectrum covariance matrix can be
obtained by integrating Eq.~(\ref{eq:pseudocov_fourI}). We propose an algorithm
that performs the computation in $\mathcal{O}(\ell_\mx ^5)$, typically gaining a
thousandfold speed-up compared to the direct implementation in
$\mathcal{O}(\ell_\mx ^6)$. This is achieved with the fast harmonic transform
tools implemented in the \healpix
library\footnote{\url{https://healpix.sourceforge.io}}. It enables the exact
computation of the covariance matrix, albeit on a limited range of multipoles.
In this work, we have computed the full covariance up to $\lexact\equiv1000$,
and calculated a few ranks of the matrix at $\ell>\lexact$. This allows the
direct comparison of the various analytic covariance approximation formulae with
the exact calculation.

In the following, we describe the algorithm we developed to perform this
computation. We validate it with Monte Carlo estimates of the covariance for the
reference \spt survey.

\subsection{Algorithm description}
\label{sec:exactTT}
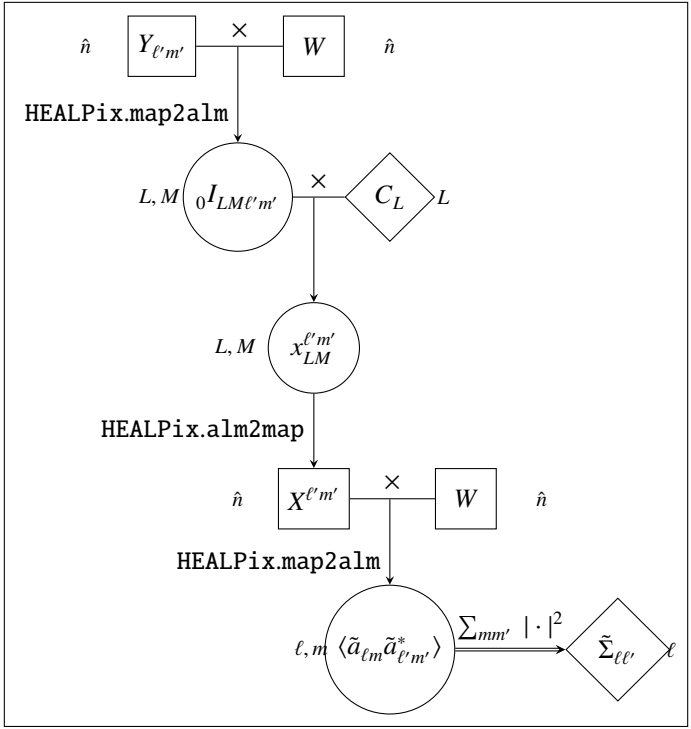
\begin{figure}[!b]
    \centering
    \begin{tikzpicture}[framed]
        \node[inner sep=0, minimum size=0 ] (inv1) at (0, 8){}; % invisible node
        \node[draw, minimum size=0.8cm, left of=inv1 ] (Y) {$Y_{\ell'm'}$};
        \node[inner sep=0, minimum size=0 , left of=Y]() {\tiny $\hat{n}$};
        % next
        \node[draw, minimum size=0.8cm, right of=inv1 ] (W1) {$W$}; \node[inner
            sep=0, minimum size=0 , right of=W1]() {\tiny $\hat{n}$};
        % next
        \node[draw, circle, minimum size=1cm ] (I) at (0,
        6){${}_{0}I_{LM\ell'm'}$}; \node[inner sep=0, minimum size=0 , left
        of=I]() {\tiny $L,M \ $};
        \node[inner sep=0, minimum size=0]
        (inv3) at (1, 6) {}; % invisible node 
        \node[draw, diamond] (C) at (2, 6) {$C_L$}; \node[inner sep=0, minimum
            size=0]() at (2.7, 6) {\tiny $L$};
        % next
        \node[draw, circle, minimum size=1.2cm ] (x) at (1, 4){$x_{LM}^{\ell'
                        m'}$}; \node[inner sep=0, minimum size=0 , left of=x]()
                        {\tiny $L,M \ $};
        % next
        \node[inner sep=0, minimum size=0 ] (inv2) at (2, 2){}; % invisible node
        \node[draw, minimum size=0.8cm, left of=inv2 ] (X) {$X^{\ell' m'}$};
        \node[inner sep=0, minimum size=0 , left of=X]() {\tiny $\hat{n}$};
        % next
        \node[draw, minimum size=0.8cm, right of=inv2 ] (W2) {$W$}; \node[inner
            sep=0, minimum size=0 , right of=W2]() {\tiny $\hat{n}$};
        % next
        \node[draw, circle, minimum size=1.5cm ] (2pt) at (2, 0) {$\langle
                \ta_{\ell m} \ta^*_{\ell ' m'} \rangle$}; \node[inner sep=0,
                minimum size=0 , left of=2pt]() {\tiny $\ell,m \ $};

        % final node
        \node[draw, diamond, minimum size=0.8cm ] (cov) at (5, 0) {
            $\tSig_{\ell\ell'}$}; \node[inner sep=0, minimum size=0] at (5.7, 0)
            () {\tiny $\ell$};

        % arrows
        \draw[-] (Y.east) -- (W1.west) node[midway,above]{$\times$};
        \draw[-stealth] (inv1.south) -- (I.north) node[midway,left, anchor=north
        east]{\healpix.\maptoalm}; \draw[-stealth] (inv3.south) -- (x.north)
        node[midway,left]{}; \draw[-] (I.east) -- (C.west)
        node[midway,above]{$\times$};\draw[-stealth] (x.south) -- (X.north)
        node[midway,left]{\healpix.\almtomap}; \draw[-] (X.east) -- (W2.west)
        node[midway,above]{$\times$}; \draw[-stealth] (inv2.south) --
        (2pt.north) node[midway,left, anchor=north east]{\healpix.\maptoalm};
        \draw[double, -stealth] (2pt.east) -- (cov.west)
        node[midway,above]{$\sum_{m m'} \ |\cdot|^2$};
    \end{tikzpicture}
    \caption{Diagram showing the algorithm to compute one row of the covariance
    $\tSig_{\ell\ell'}$ using the \healpix tools for a fixed $\ell'$ and varying
    $\ell$. Square, diamond or circle boxes are arrays representing maps, power
    spectra or spherical harmonic coefficients respectively. Operations are
    symbolized by arrows and described alongside. The indices of the arrays are
    indicated on the side of the corresponding boxes. For example, near the top
    of the diagram, the \texttt{HEALPIX.map2alm} operation applied on the
    product $Y_{l'm'}W$ produces the array $_0I_{LMl'm'}$ with indices $L,M$. At
    the bottom of the diagram, the operations before the final summation produce
    the array $\langle\ta_{\ell m} \ta^*_{\ell ' m'}\rangle$ with indices
    $\ell,m$ for a fixed $\ell',m'$ pair. This part of the algorithm scales as
    $\mathcal{O}(\ell'^3)$, since this is the scaling of \healpix operations
    (which we apply 3 times) when the resolution of the map is comparable to the
    maximum multipole index considered. The last operation, summing over the
    indices $m,m'$, requires repeating the precedent steps for all $m' \in
    [-\ell', \ell']$, thus repeating them $2\ell'+1$ times. Therefore, the final
    complexity to produce a single row with fixed multipole index $\ell'$ is
    $\mathcal{O}(\ell'^4)$. Computing the covariance matrix for all $\ell'$ then
    increases the computational time to $\mathcal{O}(\ell'^5)$. }
    \label{fig:diag}
\end{figure}

We focus on the computation of a  given row of the covariance matrix $\tSig$.
This will allow us either to compute a full covariance matrix at low multipoles,
or to test our approximations on a selection of rows. We first start by
describing the computation of the covariance of the intensity spectrum. A
diagrammatic implementation of our calculation is presented in
Fig.~\ref{fig:diag}. For the polarization spectra, the calculation follows a
similar pattern, and the difference between the two cases will be discussed in
the next section.

We describe in App.~\ref{app:covar} the derivation of the expression of the
covariance of the pseudo-spectrum leading to Eq.~(\ref{eq:pseudocov_fourI}). In
particular, we express the covariance as a sum over $m$ and $m'$ of the square
of the correlation $\langle \ta_{\ell m} \ta^*_{\ell ' m'} \rangle$
\begin{align}
    \tSig_{\ell \ell'} & = \frac{2}{(2\ell+1)(2\ell'+1)} \sum_{mm'}
    |\langle \ta_{\ell
        m} \ta^*_{\ell ' m'} \rangle|^2.
    \label{eq:defpseudocov}
\end{align}
The harmonic coefficients correlation can be written as
\begin{align}
    \langle \ta_{\ell m} \ta^*_{\ell ' m'} \rangle & = \sum_{LM} C_L \
    {}_0I_{\ell mLM}\ {}_0I^*_{\ell'm'LM},                         \nonumber                     \\
                                                   & = \int \ud \hu \ {}_0Y_{\ell m}(\hu) W(\hu)
    \sum_{LM} \left\{ C_L \ {}_0I^*_{\ell'm'LM}\right\}\  {}_0Y^*_{LM}(\hu),
    \label{eq:laststep2pt_1}
\end{align}
where we have used Eq.~(\ref{eq:polIkernels}) to expand one of the ${}_0I$
kernels, re-organized the equation, and used the fact that the power spectrum
$C_L$ and the mask $W$ are real quantities (we also dropped the explicit $W$
dependencies of the kernel to simplify notations). For fixed $\ell'$ and $m'$,
the rightmost part of the equation can be seen as the complex conjugate of the
backward spherical harmonic transform of a set of spherical harmonic
coefficients into a map $X^{\ell'm'}$, defined as
\begin{align}
    X^{\ell'm'}(\hu) \equiv \sum_{LM} x^{\ell'm'}_{LM}Y_{LM}(\hu),
    \label{eq:mapX}
\end{align}
where we defined the spherical harmonic coefficients with
\begin{align}
    x^{\ell'm'}_{LM} \equiv C_L \  {}_0I^*_{LM\ell'm'}.
    \label{eq:definex}
\end{align}
Here, we emphasize that the map $X^{\ell'm'}$ is a complex map, thus it needs
special care when decomposing into harmonic coefficients. With these
definitions, Eq.~(\ref{eq:laststep2pt_1}) reduces to
\begin{align}
    \langle \ta_{\ell m} \ta^*_{\ell ' m'} \rangle
     & =  \int \ud \hu \ {}_0Y_{\ell m}(\hu) W(\hu) X^{\ell'm'*}(\hu),
    \label{eq:laststep2pt}
\end{align}
where we recognize the forward harmonic transform of the map $X^{\ell'm'}$,
masked by $W$. Thus, a spherical harmonic transform of a masked map can produce
the correlation $\langle \ta_{\ell m} \ta^*_{\ell ' m'} \rangle $ for all
$\ell,m$ and a fixed pair of $\ell',m'$. As we discussed, this $X^{\ell'm'}$ map
is defined by a set of spherical harmonic coefficients whose expression is given
in Eq.~(\ref{eq:definex}).

The computation of the $x^{\ell'm'}_{LM}$ coefficients requires the evaluation
of the ${}_0I$ kernel. Using Eq.~(\ref{eq:polIkernels}) for a fixed $\ell',m'$,
the ${}_0I_{LM\ell'm'}$ kernel can be computed as the spherical harmonic
transform of a masked ${}_0Y_{\ell' m'}$ map, for all of the $L,M$ indices.
Putting everything together, we see that for a choice of $\ell',m'$, the
computation of $\langle \ta_{\ell m} \ta^*_{\ell ' m'} \rangle $ for all $\ell,
m$ reduces to two forward and one backward spherical harmonic transforms, as
summarized in Fig.~\ref{fig:diag}.

In practice, these decompositions can be performed with \healpix, which takes
advantage of a specific pixelation scheme to make the computation more
efficient. This is where the gain announced at the beginning of this section
comes from and allows us to implement the exact computation. \healpix
decompositions typically scale as $\mathcal{O}(\ell'^3)$\footnote{Details about
the computation scaling of \healpix can be found on the website or in
\cite{Gorski2005}}. We repeat the decompositions resulting in $\langle \ta_{\ell
m} \ta^*_{\ell ' m'} \rangle$ for all $m' \in [-\ell',\ell']$ to perform the
summation in Eq.~(\ref{eq:defpseudocov}). Thus, the computation of a single rank
$\tSig_{\ell\ell'}$ for all $\ell$ and fixed $\ell'$  scales as
$\mathcal{O}(\ell'^4)$. Finally, the computation of a full covariance matrix for
all $\ell'$ scales as $\mathcal{O}(\lexact^5)$.

Additional optimisations can be implemented in the algorithm, namely by degrading
maps and running the algorithm at a lower \healpix resolution, \nside, for small
multipoles. \healpix computations are precise up to $\ell \sim 2\nside$, hence
choosing a map resolution of the order of the multipole is sufficient to compute
precisely the close-to-diagonal elements of the covariance. This operation
requires a degraded version of the mask, which must be computed while avoiding
aliasing from small scale features. This can be done by implementing a hard
cut-off of the mask harmonic coefficients before degrading its resolution. Doing
so allowed us to compute the exact covariance up to multipole $\lexact=1000$.
The algorithm requires 300h of \textsc{CPU}-time to compute a row of the
intensity (\TTTT) and polarization (\EEEE) matrices at multipole $\ell=950$ with
map resolution $\nside=1024$. It is also well suited to a potential GPU
implementation, which could lead to more speed-ups.

\subsection{Polarization}
\label{sec:exactEE}
The polarized case is very similar to the intensity one detailed in the previous
subsection. We only describe the \EEEE case which gives a general template to
the other polarization and temperature$\times$polarization cases.

The polarized version of Eq.~(\ref{eq:laststep2pt_1}) is given by Eq.~(6) of
\cite{Challinor2004a} that writes
\begin{align}
    \langle \ta^\E_{\ell
        m} \ta^{\E*}_{\ell ' m'} \rangle = \nonumber
    \sum_{LM} \big[ & C^{\EE}_L {}_+I_{\ell mLM} {}_+I_{\ell'm'LM}^*
    +                                                                \\ &C^{\BB}_L {}_-I_{\ell mLM} {}_-I_{\ell'm'LM}^* \big],
\end{align}
where we defined the Hermitian coupling coefficients
\begin{equation}
    {}_{\pm}I_{\ell mLM} = \frac{1}{2} \left ( {}_{+2}I_{\ell mLM}
    \pm {}_{-2}I_{\ell mLM} \right),
\end{equation}
with the spin-weighted coupling coefficients ${}_{\pm2}I$ defined as for
intensity, see Eq.~(\ref{eq:polIkernels}). Reordering the terms and doing the
same operations as in the previous section, the final harmonic coefficient
correlation can be seen as the masked forward spherical harmonic decomposition
of two maps $Z^{\ell'm'}_1, Z^{\ell'm'}_2$,
\begin{align}
    \label{eq:2pteewithZ}
    \langle \ta^\E_{\ell
        m} \ta^{\E*}_{\ell ' m'} \rangle =
    \frac{1}{2} \left[ \int \mathrm{d} \hu W(\hu)Z^{\ell'm'}_1(\hu)
        \left ( {}_{+2}Y^*_{\ell
            m} + {}_{-2}Y^*_{\ell
    m} \right ) (\hu) \right. \\
        \nonumber \left.
        - i \int \mathrm{d} \hu W(\hu)Z^{\ell'm'}_2(\hu)
        \left ( {}_{+2}Y^*_{\ell
            m} - {}_{-2}Y^*_{\ell
            m} \right )(\hu) \right].
\end{align}
The maps $Z^{\ell'm'}_1, Z^{\ell'm'}_2$ are obtained using a backward spherical
harmonic decomposition of the coefficients $x_{LM}^{\ell'm';\E,\B}$,
\begin{align}
    (Z^{\ell'm'}_1-iZ^{\ell'm'}_2)(\hu) & \equiv \sum_{LM} \frac{x^{\ell'm';\E}_{LM} +
    x^{\ell'm';\B}_{LM}}{2} {}_{+2}Y_{LM}(\hu),                                         \\
    (Z^{\ell'm'}_1+iZ^{\ell'm'}_2)(\hu) & \equiv  \sum_{LM} \frac{x^{\ell'm';\E}_{LM} -
        x^{\ell'm';\B}_{LM}}{2} {}_{-2}Y_{LM}(\hu).
\end{align}
The set of harmonic coefficients $x_{LM}^{\E,\B}$ are defined similarly to the
temperature case (Eq.~(\ref{eq:definex})), and obtained by filtering the
coefficients computed with a masked forward harmonic decomposition of the spin-2
spherical harmonics ${}_{\pm2}Y_{\ell'm'}$,
\begin{equation}
    \left\{
    \begin{matrix}
        x^{\ell'm';\E}_{LM}  = C_L^{\EE} {}_+I_{\ell'm'LM}, \\
        x^{\ell'm';\B}_{LM}  = C_L^{\BB} {}_-I_{\ell'm'LM}.
    \end{matrix}
    \right.
\end{equation}
This algorithm can be extended for any combination of spectra for the other
polarization cases, including the cross-correlation between temperature and
polarization that we don't treat here.

\subsection{Validation on simulations}

We compare the results of our implementation of the exact computation with a
Monte-Carlo estimate of the covariance, obtained with \Nsim simulations. The
Monte-Carlo covariance terms are expected to be Wishart distributed with \Nsim
degrees of freedom, as explained in \cite{Lueker2010}. We can estimate their
variance to be
\begin{equation}
    \label{eq:statvarofmccov}
    \left\langle \left( \tSig_{\ell\ell'}^\sims -
    \left\langle \tSig_{\ell\ell'}^\sims
    \right\rangle \right)^2\right\rangle
    =
    \frac{\tSig_{\ell\ell'}^2 +
        \tSig_{\ell\ell}\tSig_{\ell'\ell'}}{\Nsim}.
\end{equation}

$\Nsim=10\, 000$ allows us to reach a percent level accuracy on the diagonal.
This is the number of realizations that we will use for the Monte-Carlo
covariance. For this validation, we will use the mask shown in
Fig.~\ref{fig:maskonsphere}. Note that in this idealized setting, we do not
include a point source mask.

We perform an exact computation of the \TTTT and \EEEE covariance up to
$\lexact=1000$, using our algorithm and degrading the mask to smaller
resolutions. Furthermore, we compute the rows every 25 multipoles of the matrix
up to $\lmax=1500$, as well as at a few well chosen multipoles, corresponding to
peaks and troughs of the spectra, up to $\lmax=2000$.
\begin{figure}[!ht]
    \centering
    \includegraphics{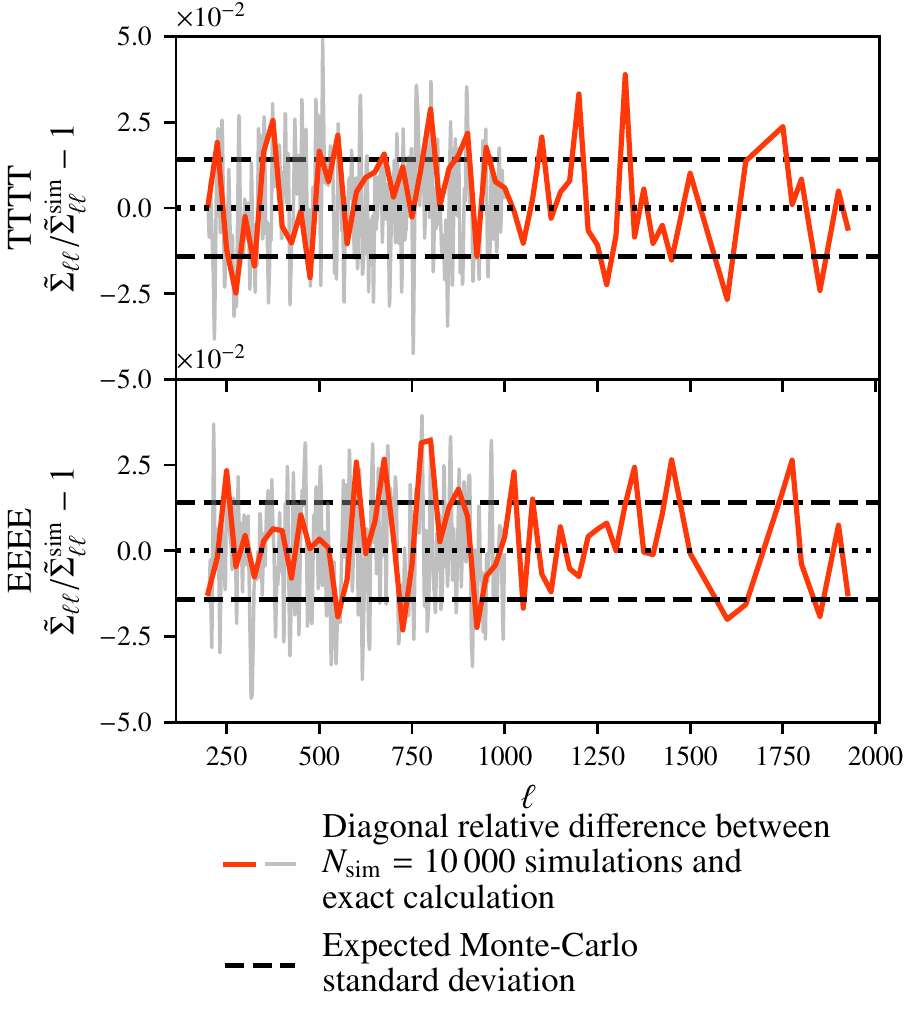}
    \caption{Relative difference of diagonals $\tSig_{\ell\ell}/\tSig^
            {\sims}_{\ell\ell} -1$ for temperature \TTTT (top) and polarization
            \EEEE (bottom). In red are the relative differences \emph{every 25
            multipoles} until $\ell = 1500$ and a few well chosen ones (at the
            locations of peaks and troughs of the spectra) up to $\ell=2000$. In
            grey, the same quantity is plotted \emph{for all multipoles} for
            $\ell \in [\lcut=200, \lexact=1000]$. We are able to compute exactly
            the covariance only for a limited number of rows, and it is
            computationally cheaper for lower multipoles, justifying our choice
            of full calculation at $\ell<\lexact$ and partial calculation for
            larger multipoles. This plot shows the agreement between the two
            approaches, validating our exact calculation.}
    \label{fig:simsvsexact}
\end{figure}

\begin{figure*}[!ht]
    \centering
    \includegraphics{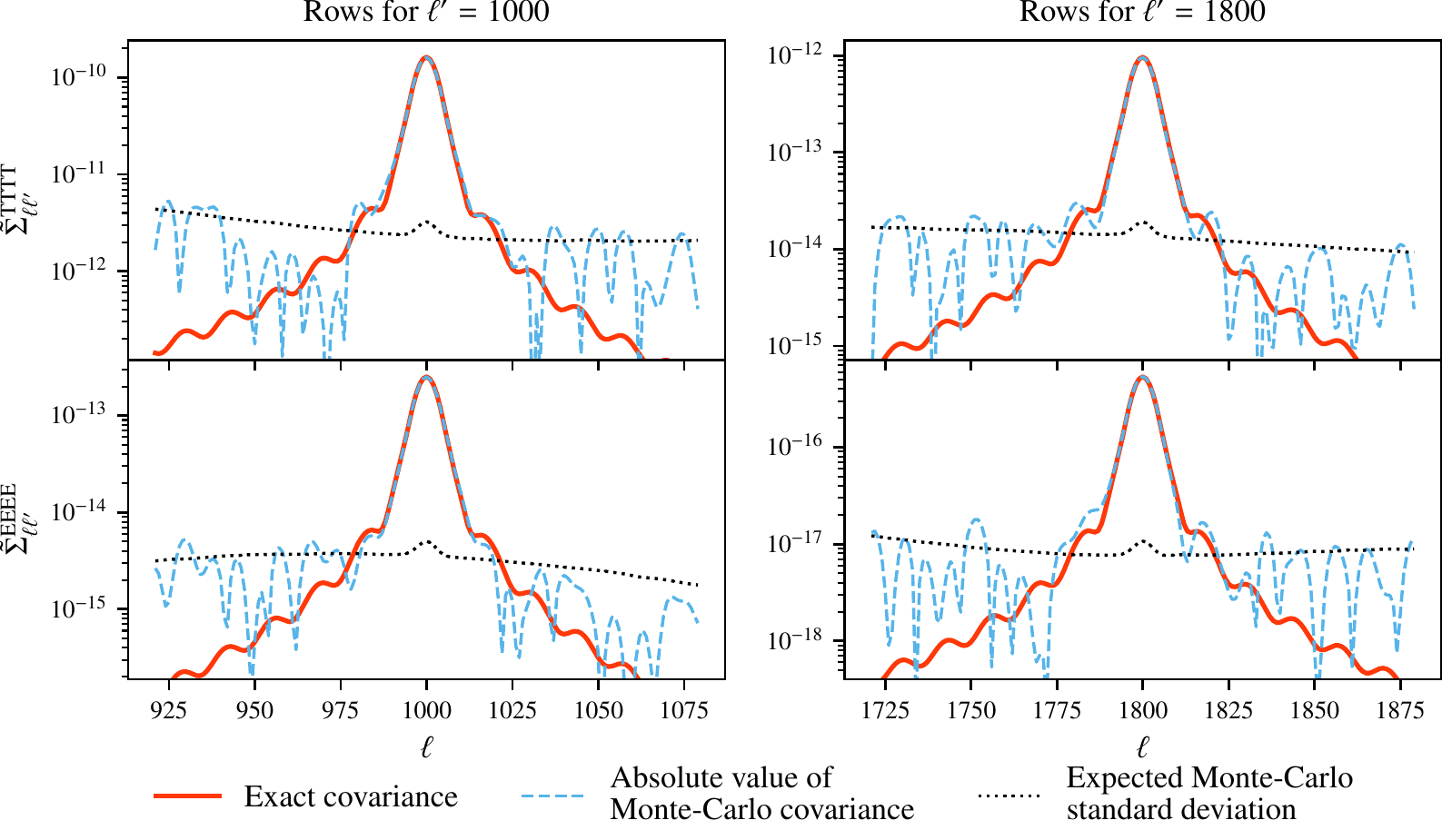}
    \caption{Rows of the exact covariance (solid red) and of the simulated one
        (dash blue), for multipole $\ell'=1000$ (\lhs) and $\ell'=1800$ (\rhs).
        The dotted black line shows the expected Monte-Carlo uncertainty on the
        simulated covariance from Eq.~(\ref{eq:statvarofmccov}), using
        $\Nsim=10\, 000$ realizations. There is excellent agreement between the
        two approaches. The Monte-Carlo standard deviation is as large as the
        covariance off-diagonal terms at $|\ell-\ell'|\sim25$.}
    \label{fig:rowsimsvsexact}
\end{figure*}

We first focus on the diagonal of the covariance. Figure \ref{fig:simsvsexact}
presents the comparison between the exact computation of the diagonal, obtained
by selecting the corresponding value in the rows we have computed, and the MC
evaluation. The two agree within the MC noise expected for $\Nsim$.

\begin{figure}[!ht]
    \centering
    \includegraphics{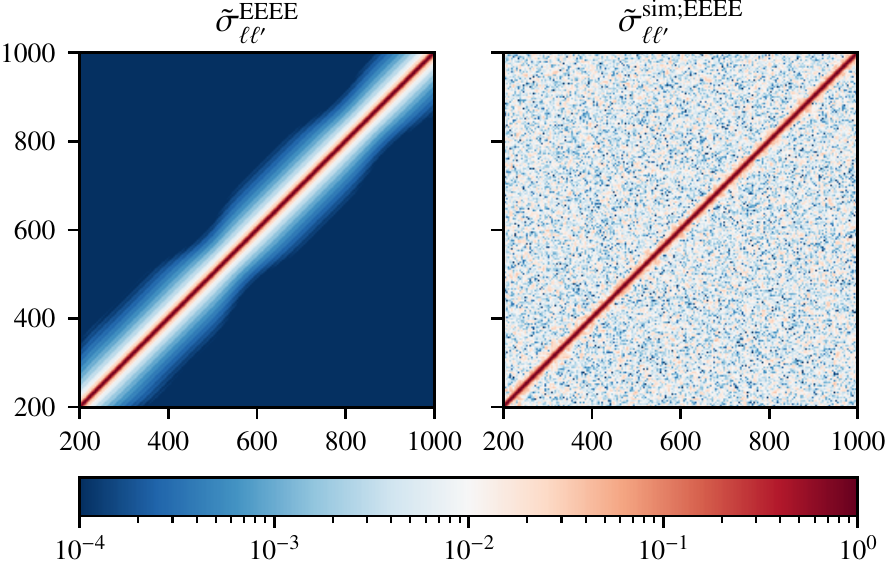}
    \caption{\EEEE correlation matrix (see Eq.~(\ref{eq:covtocorr})) obtained
        from the exact (left) and Monte-Carlo (right) calculations of the
        covariance matrix. The exact computation allows us to have the full
        correlation matrix, whilst the Monte-Carlo approach is limited by
        numerical noise. All terms below $10^{-4}$ are plotted in dark blue.}
    \label{fig:2dcov}
\end{figure}

For what concerns the off-diagonal terms, we show in
Fig.~\ref{fig:rowsimsvsexact} a few rows of the exact and Monte-Carlo
covariance, which agree within MC noise. The correlation between modes falls to
the percent level within a distance $|\ell-\ell'|\sim 25$ band around the
diagonal. The correlation matrix is defined as the covariance renormalised by
its diagonal
\begin{align}
    \sigma_{\ell\ell'} \equiv \frac{\Sigma_{\ell\ell'}}{\sqrt{\Sigma_{\ell\ell}
            \Sigma_{\ell'\ell'}}}.
    \label{eq:covtocorr}
\end{align}
We display the exact and Monte-Carlo correlation matrix on the same multipole
range in Fig.~\ref{fig:2dcov}.

While massive Monte-Carlo estimates, as we performed here in this idealized
case, can produce accurate off diagonal terms estimates, performing a large
number of Monte-Carlo can be more challenging in the case of a realistic
experiment, requiring regularization approaches, where the number of possible
simulations is limited by the computational cost of mock-observations.

We also stress that our covariance matrix cannot be directly compared to the one
used in \cite{Dutcher2021}, since the presence of a point source mask (that we
did not include in our simple example), the complications brought about by
introducing realistic noise and scanning strategy, projection effects (the
analysis in \cite{Dutcher2021} is performed using a flat sky approach) can all
yield to different levels of correlations between modes.

Our tests demonstrate that our implementation of the exact computation of the
pseudo-spectrum covariance is correct, at least at the level of accuracy that
can be reached by Monte-Carlo estimation.

%%%%%%%%%%%%%%%%%%%%%%

%%%%%%%%%%%%%%%%%%%%%%
\section{Existing approximations and their accuracy on a small patch of the sky}
\label{sec:oldapprox}
Our algorithm allows us to obtain the exact covariance only for $\ell<1000$ or
for a few rows at higher $\ell's$, due to the expensive computing resources
required. The usual analytical approach consists in using approximations of
Eq.~(\ref{eq:pseudocov_fourI}) to decrease the computational cost. In this
section, we introduce a new framework to express the approximations of the
covariance matrix and use it to list the different methods proposed in the
literature. Then, we test and discuss their accuracy against our exact
computation.

\subsection{General framework}
Before discussing the approximations of the covariance, we define a few
quantities that help relate the various approximations to each other. We rewrite
Eq.~(\ref{eq:pseudocov_fourI}) as
\begin{equation}
    \label{eq:covwithX}
    \tSig_{\ell\ell'} = \frac{2}{(2\ell+1)(2\ell'+1)}
    \sum_{\ell_1\ell_2}
    C_{\ell_1} \Theta^{\ell_1\ell_2}_{\ell\ell'}[W] C_{\ell_2},
\end{equation}
introducing $\Theta^{\ell_1\ell_2}_{\ell\ell'}[W]$,  the \emph{covariance
    coupling kernel}, defined as the sum over the multipole orders ($m,
    m',\dots$) of the coupling coefficients reads
\begin{equation}
    \label{eq:deffourcX}
    \Theta^{\ell_1\ell_2}_{\ell\ell'}[W] \equiv \sum_{mm_1m'm_2}
    ({}_{0}I_{\ell m\ell_1m_1} \ {}_{0}I_{\ell_1m_1\ell'm'} \
    {}_{0}I_{\ell'm'\ell_2m_2} \ {}_{0}I_{\ell_2m_2\ell m})[W].
\end{equation}
The covariance coupling kernel $\Theta$ represents the coupling between the
modes of the theoretical underlying power spectrum $C_{\ell_i}$ for $i=1,2$
depending on the index of the pseudo-covariance $(\ell,\ell')$. We chose to show
the two indices related to the covariance $(\ell,\ell')$ as subscripts, and the
two summing indices $(\ell_1,\ell_2)$ as superscripts. In this work, we consider
a single-mask temperature case, for which the coupling kernel is symmetric with
respect to the exchange of multipole indices $\ell \leftrightarrow \ell'$ or
$\ell_1 \leftrightarrow \ell_2$. In the following, we will write our results in
this case for the sake of simplicity, but they are valid regardless of the
choice of single or multiple masks. In the case of spectra obtained from maps
with different masks, or the case of cross-spectra, the kernel is not symmetric.
While the results of this work apply to both cases, considering multiple masks
increases computing cost.

Using the closing relations of spherical harmonics, given in
Eqs.~(\ref{eq:relationofI1}) and (\ref{eq:relationofI2}), we can write,
\begin{equation}
    \label{eq:norm_fourcX}
    \sum_{\ell_1\ell_2} \Theta^{\ell_1\ell_2}_{\ell\ell'}[W]
    = (2\ell+1)(2\ell'+1)\Xi^{00}_{\ell\ell'}[W^2].
\end{equation}
We can now define the \emph{reduced covariance coupling kernel} as
\begin{equation}
    \bThe_{\ell\ell'}^{\ell_1\ell_2}[W] \equiv
    \frac{\Theta^{\ell_1\ell_2}_{\ell\ell'}[W]}
    {(2\ell+1)(2\ell'+1)\Xi^{00}_{\ell\ell'}[W^2]},
\end{equation}
for which
\begin{equation}
    \label{eq:norm_fourcbarX}
    \sum_{\ell_1\ell_2}
    \bThe_{\ell\ell'}^{\ell_1\ell_2}[W] = 1.
\end{equation}
With these notations, one can rewrite the covariance as
\begin{equation}
    \label{eq:covwithbarX}
    \tSig_{\ell\ell'} = 2
    \Xi^{00}_{\ell\ell'}[W^2]
    \sum_{\ell_1\ell_2}
    C_{\ell_1} \bThe_{\ell\ell'}^{\ell_1\ell_2}[W] C_{\ell_2}.
\end{equation}
The symmetric mode-coupling kernel $\Xi[W^2]$ provides the purely geometric
coupling due to sky masking and is common to all approximations of the
covariance. It only depends on the power spectrum of the squared mask. Its
computation scales as $\mathcal{O}(\lmax^3)$. This could be improved by noting,
as was done by \cite{Louis2020}, that at small enough scales, $\Xi[W^2]$ is
close to a Toeplitz matrix allowing us to further reduce the scaling to
$\mathcal{O}(\lmax ^2)$ for a large range of modes. The sum on the right
hand-side of Eq.~(\ref{eq:covwithbarX}) describes the contribution of the signal
power spectrum modulated by the kernel $\bThe$, which depends on the mask. It is
this sum that all approximations try to simplify, replacing the kernel $\bThe$
with a simpler ansatz. In the following, we will describe all approximations in
terms of this redefinition of the covariance matrix.

\subsection{Approximations}
\subsubsection{NKA}
Based on the observation that the coupling coefficients ${}_{0}I$ in
Eq.~(\ref{eq:tildea_I_alm}) are narrow and peaking at their first multipole
indices $\ell$ or $\ell'$, \cite{Efstathiou:2004} introduced the following
approximation of Eq.~(\ref{eq:covwithbarX}), taking the convolving spectra
$C_{\ell_i}$, $i=1,2$ out of the sum, and replacing them by the power spectrum
evaluated at the first multipole index of the coupling coefficients (\ie the
covariance indices of $\bThe$), $C_\ell$ or $C_{\ell'}$. Following the notation
introduced in \cite{Garcia-Garcia2019}, we will refer to this approximation of
the covariance as \NKA (Narrow Kernel Approximation).
\begin{align}
    \tSig_{\ell \ell'} \approx & \ 2C_\ell C_{\ell '} \Xi^{00}_{\ell\ell'}[W^2]
    \sum_{\ell_1\ell_2} \bThe_{\ell\ell'}^{\ell_1\ell_2}[W] \nonumber                       \\
    =                          & \  2C_\ell C_{\ell '} \Xi^{00}_{\ell\ell'}[W^2] \ \equiv \
    \tSig^\NKA_{\ell \ell'}. \label{eq:NKA}
\end{align}
In terms of the reduced covariance coupling kernel, the \NKA approximation uses
\begin{equation}
    \bThe_{\ell\ell'}^{\ell_1\ell_2}[W]
    \approx \bThe_{\ell\ell'}^{\ell_1\ell_2;\NKA}[W]
    \equiv
    \frac{\delta_{\ell\ell_1}\delta_{\ell'\ell_2} +
        \delta_{\ell'\ell_1}\delta_{\ell\ell_2}}{2}.
    \label{eq:NKAThetaform}
\end{equation}
The approximation is exact for the full sky or for a constant underlying power
spectrum $C_\ell = N$. It provides an accurate estimator whenever the underlying
power spectrum $C_{\ell}$ varies slowly as a function of $\ell$ compared to
the typical size of the operators ${}_{0}I$. This condition is fulfilled when
the amplitude of the mask power spectrum drops quickly with multipole $\ell$,
which is the case for large sky fractions observed with a mask which contains no
small scale features. This is shown for example in Fig.~\ref{fig:maskonsphere},
where we plot the power spectrum of one of the masks used in the \planck
analysis. In this case, the above approximation holds for multipoles much larger
than those for which the mask spectrum contains power.

The \NKA approximation was first introduced in intensity by
\cite{Efstathiou:2004} and extended to polarization in \cite{Challinor2004a}. As
in the temperature case, the approximated covariances in polarization are
expressed as a function of the polarization spectra \EE and \BB and the
symmetric coupling kernels $\Xi_{\ell,\ell'}^{\pm2,\pm2}$. The expressions of
the approximated polarization covariances mix \EE and \BB due to leakage which
appears because the sky is masked, see Eq.~(25-27) of \cite{Challinor2004a}.

The \NKA approximation has been widely used, for instance in the \planck
cosmological analysis, which masked only small portions of the full sky, see
\cite{PlanckCollaboration2015}. However, it has never been thoroughly tested on
small sky fractions. As shown in Fig.~\ref{fig:maskonsphere}, the mask power
spectrum in the case of the small survey footprint of \spt drops much slower
than the large \planck one. From this observation, we expect the mode coupling
kernels ${}_sI$ to be wider, as it can be deduced from Eq.~\ref{eq:polIkernels}.
As a result, the theoretical underlying spectrum $C_\ell$ might not be treated
as constant compared to the covariance coupling kernels in the sums of
Eq.~(\ref{eq:covwithbarX}), and \sptyeete may be outside the regime of validity
of the \NKA assumption. This will be tested at the end of this section. We will
now list some proposed improvements to the \NKA approximation.

\subsubsection{FRI}
A straightforward extension of the \NKA approximation has been proposed in
\cite{Friedrich2021}. It is based on the observation that the reduced covariance
coupling kernel $\bThe$ has four maxima at $[ \ell=\ell_1, \ell'=\ell_1]$,
$[\ell=\ell_1, \ell'=\ell_2]$, $[\ell=\ell_2, \ell'=\ell_1]$ and $[\ell=\ell_2,
\ell'=\ell_2]$. This suggests the following form of the reduced covariance
coupling matrix:
\begin{equation}
    \bThe_{\ell\ell'}^{\ell_1\ell_2}[W]
    \approx \bThe_{\ell\ell'}^{\ell_1\ell_2;\FRI}[W]
    \equiv
    \frac{\delta_{\ell\ell_1} + \delta_{\ell'\ell_1}}{2}
    \frac{\delta_{\ell\ell_2} + \delta_{\ell'\ell_2}}{2}.
\end{equation}
Thus, the approximated covariance is
\begin{equation}
    \label{eq:FRI}
    \tSig_{\ell\ell'}^\FRI
    \equiv 2
    \Xi^{00}_{\ell\ell'}[W^2]
    \left(\frac{C_\ell + C_{\ell'}}{2}\right)^2.
\end{equation}
We will refer to this approximation as \FRI in the rest of the article.

\subsubsection{INKA}
\cite{Nicola2021} proposed an improved version of the \NKA, the \INKA (Improved
Narrow Kernel Approximation). In this approximation, the Dirac functions in
Eq.~(\ref{eq:NKAThetaform}) are replaced by ${}_{0}\bar{M}$, the renormalised
\master mode-coupling kernel, as defined in App.~\ref{sec:apprenormalised}. It
writes
\begin{equation}
    \label{eq:definka}
    \bThe_{\ell\ell'}^{\ell_1\ell_2}[W]
    \approx \bThe_{\ell\ell'}^{\ell_1\ell_2;\INKA}[W]
    \equiv
    \frac{{}_{0}\bar{M}_{\ell\ell_1}{}_{0}\bar{M}_{\ell'\ell_2} +
    {}_{0}\bar{M}_{\ell'\ell_1}{}_{0}\bar{M}_{\ell\ell_2}}{2}.
\end{equation}
Indeed, the convolution in Eq.~(\ref{eq:pseudocov_fourI}) averages the power
spectra $C_{\ell_i}, i=1,2$ over multipoles close to $\ell$ and $\ell'$. One can
take advantage of this by replacing the convolution by a multiplication with a
smoothed power spectrum. If one defines $\bar{C} \equiv \bar{M}C$, the resulting
covariance can be written as
\begin{align}
    \label{eq:INKA}
    \tSig_{\ell\ell'}^\INKA
    \equiv 2 \bar{C}_\ell \bar{C}_{\ell'}
    \Xi^{00}_{\ell\ell'}[W^2].
\end{align}

All the \NKA, \FRI and \INKA scale as $\mathcal{O}(\lmax ^3)$, which are the
computing resources needed to obtain the coupling kernels $\Xi$ and $\bar{M}$.
This is a significant improvement over the $\mathcal{O}(\lmax ^6)$ scaling of
the full computation. We will now turn to validating the approximations in the
case of small surveys, using the expensive exact computation of the covariance
matrix.

%%%
\subsection{Accuracy}
We test the accuracy of the \NKA, \FRI and \INKA approximations using the exact
computation in the case of the \spt small survey footprint shown in
Fig.~\ref{fig:maskonsphere}. For this mask, the correlations between modes are
significant, as already seen in Fig.~\ref{fig:rowsimsvsexact}. In this case, it
is customary to bin the individual multipoles into wider bandpowers. For this
reason, we will perform all of our tests on a binned version of the covariance.
Given the shape of the power spectrum of the mask and the correlations that we
expect from it, we adopt a $\Delta \ell =50$ binning with $\ell(\ell+1)/(2\pi)$
weights to flatten the dynamics of the spectra in each bin. With this bin size,
we expect that most of the correlations between bandpowers is concentrated in
the first off-diagonal bin. We will also conservatively exclude the first
$\lcut=200$ multipoles from our analysis. Those are more challenging to measure
on small survey footprint as they can suffer from leakage from the super survey
scales. We will restrict our comparison to the multipoles between $\lcut$ and
$\lexact=1000$, where we have carried out the exact computation of all of the
matrix rows.

\begin{figure*}[!ht]
    \centering
    \includegraphics{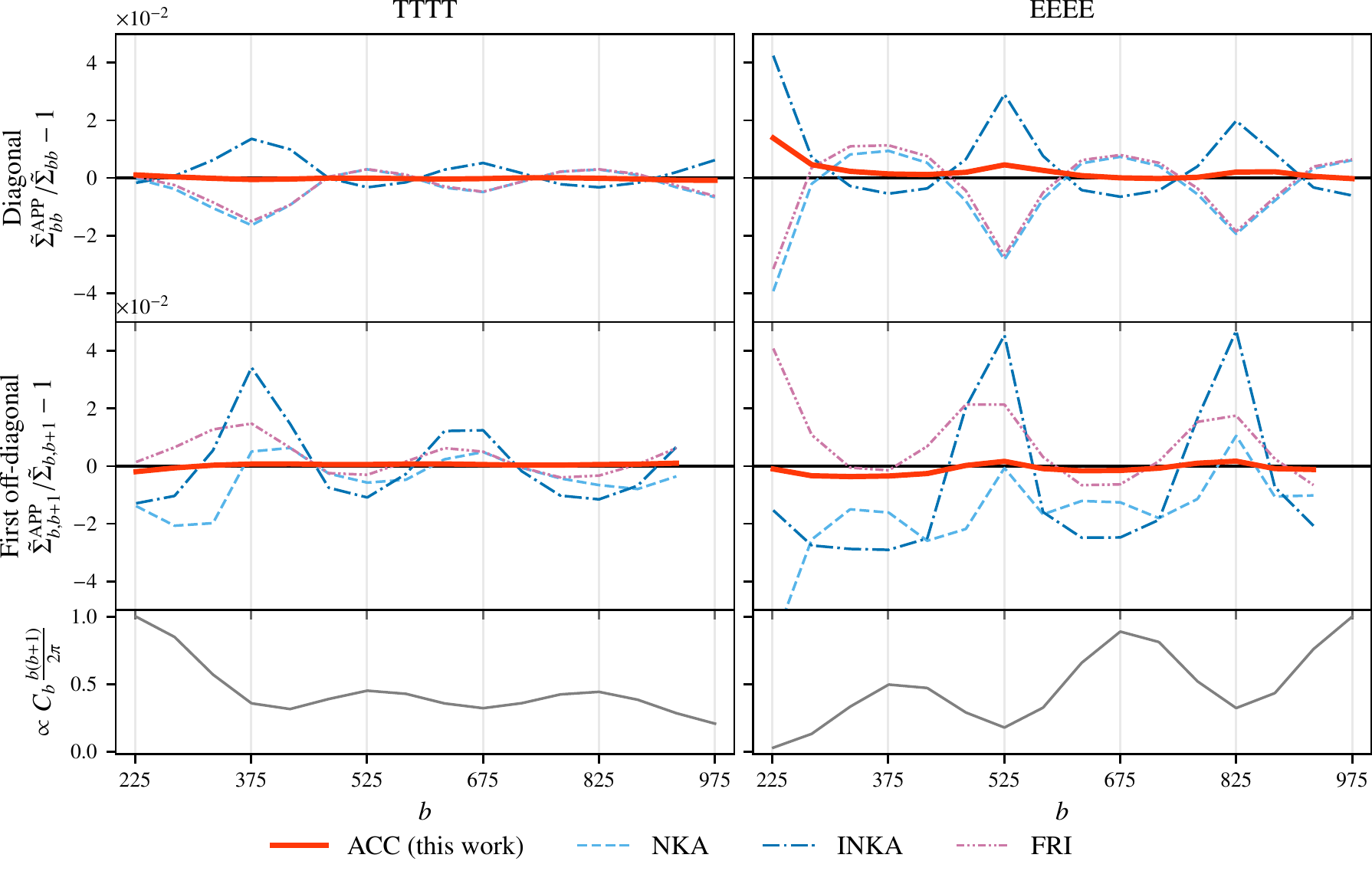}
    \caption{Relative differences of binned approximations with respect to the
        exact binned covariance: $\tSig^\mathrm{APP}_{bb'}/\tSig_{bb'} - 1$, for
        \TTTT (\lhs) and \EEEE (\rhs), with binning $\Delta \ell = 50$ . On the
        first row are plotted the relative differences for the diagonal, \ie
        $b=b'$, while on the second row are plotted those of the first
        off-diagonal, \ie $b'=b+1$. The \NKA (light blue dash), \FRI (purple
        dash-double-dot) and \INKA (dark blue dash-dot) approximations are
        accurate at the 5\% level, whereas the \ACC approximation (solid red)
        reaches the $1\%$ level, both in intensity and polarization for
        multipoles larger than $\lcut=200$. The relative differences are plotted
        for bins that include multipoles up to $\lexact=1000$ since it is the
        maximum multipole for which we have computed all the rows of the exact
        covariance. The third row displays the corresponding binned underlying
        renormalised spectrum \TT or \EE, to put forward that the difference on
        the covariances are on the peaks and troughs of the spectra, since this
        is where the correlation between adjacent scales will have the most
        impact.}
    \label{fig:approxvsexact_pseudo}
\end{figure*}

We present in Fig.~\ref{fig:approxvsexact_pseudo} a comparison between the exact
computation and the \NKA, \FRI and \INKA approximations, for the diagonal and
first off-diagonal of the \TTTT and \EEEE binned covariances. We will discuss
the performance of our new \ACC approximation, also shown in the figure, later
in the next section. The existing approximations provide good estimates of those
elements of the covariance as they fall within $5\%$ of accuracy. The amplitude
of the errors vary at different multipoles. Even though the \FRI and \INKA
schemes were implemented to improve upon the simple \NKA approximation, their
errors are of similar amplitude for this choice of binning. However, all
approximations fail to recover the binned covariance at the percent level. In
Sect.~\ref{sec:newapprox}, we will use the knowledge gained from the exact
computation to propose an improved approximation scheme.

\begin{figure}[!ht]
    \centering
    \includegraphics{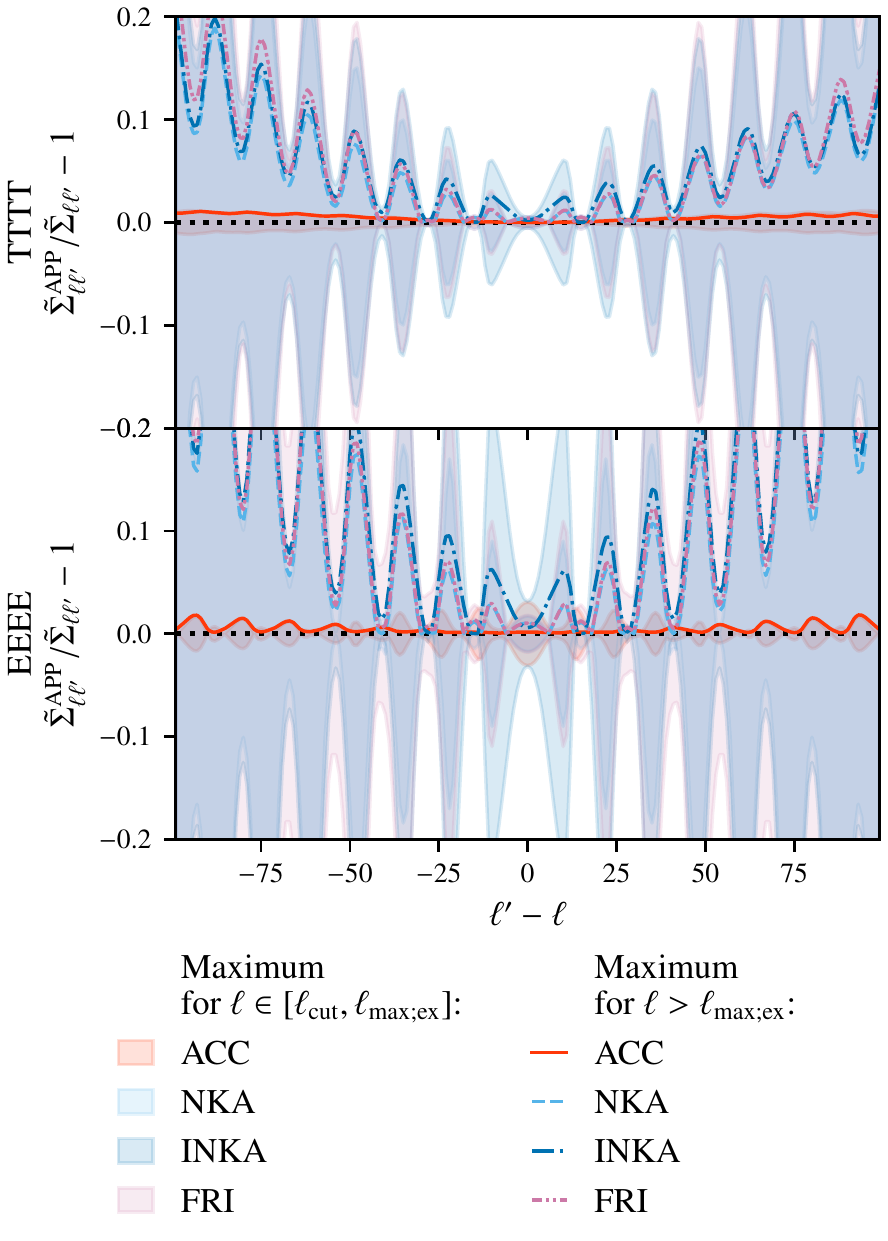}
    \caption{Relative difference between the unbinned approximated covariance
        matrices compared to the exact calculation,
        $\tSig^{\textsc{APP}}_{\ell\ell'}/\tSig_{\ell\ell'} -1$, as a function
        of $\Delta=\ell-\ell'$.      We show the \NKA (light blue), \INKA (dark
        blue) and \FRI (purple) approximations for \TTTT (top) and \EEEE
        (bottom). We only show the relative differences for the $\ell$ at which
        the deviation is largest, with $\ell\in[\lcut=200,\lexact = 1000]$
        (shaded regions) or $\ell\in[\lexact, 2000]$ (lines). By comparing the
        two, one can see that the approximations perform with a similar level of
        accuracy at low and high-$\ell$. For comparison, we also plot our new
        approximation \ACC (red), which we will introduce in
        Sect.~\ref{sec:newapprox}.The \ACC approximation improves over the
        others by a factor of $\sim 4$. }
    \label{fig:highmultipolerows_old}
\end{figure}

At higher multipoles, since we cannot easily compute the full matrix to present
binned results, we only compare some unbinned rows in
Fig.~\ref{fig:highmultipolerows_old}. The shaded regions in this figure give the
worst values of the relative difference for the approximation within multipoles
$\ell \in [\lcut=200, \lexact=1000]$ and $\ell' \in [\ell-2\Delta\ell,
\ell+2\Delta\ell]$. Those are the covariance terms for which we can calculate
the full binned covariance, and whose accuracy is shown in
Fig.~\ref{fig:approxvsexact_pseudo}. Furthermore, the lines in
Fig.~\ref{fig:highmultipolerows_old} show the same quantity as the shaded
regions, \ie the maximal relative difference, but for multipoles $\ell \in
[\lexact, 2000]$, estimated over the sparse number of rows for which we have
computed the matrix exactly. We see that the difference with the exact
covariance for all approximations at large multipoles are always within the same
error range as for lower multipoles. This shows that the approximations still
work with the same precision at higher multipoles, both for temperature and
polarization and that the accuracy of the approximations in the unbinned case
quickly falls below $20\%$ when $\Delta=50$.

%%%
\subsection{ Structure of the reduced covariance coupling kernel}

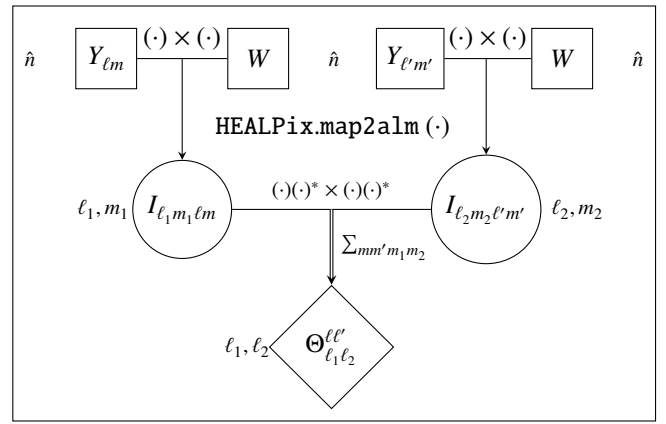
\begin{figure}[!ht]
    \centering
    \begin{tikzpicture}[framed]
        \node[inner sep=0, minimum size=0 ] (inv1) at (0, 8){}; % invisible node
        \node[draw, minimum size=0.8cm, left of=inv1 ] (Y) {$Y_{\ell m}$};
        \node[inner sep=0, minimum size=0 , left of=Y]() {\tiny $\hat{n}$};
        % next
        \node[draw, minimum size=0.8cm, right of=inv1 ] (W1) {$W$}; \node[inner
            sep=0, minimum size=0 , right of=W1]() {\tiny $\hat{n}$};
        % next
        \node[draw, circle, minimum size=1cm ] (I) at (0, 6){$I_{\ell_1 m_1\ell
                        m}$}; \node[inner sep=0, minimum size=0 , left of=I]()
                        {\tiny $\ell_1, m_1 \ $}; \node[inner sep=0, minimum
                        size=0 ] (inv2) at (4, 8){};
        % invisible node
        \node[draw, minimum size=0.8cm, left of=inv2 ] (Y2) {$Y_{\ell' m'}$};
        % next
        \node[draw, minimum size=0.8cm, right of=inv2 ] (W2) {$W$}; \node[inner
            sep=0, minimum size=0 , right of=W2]() {\tiny $\hat{n}$};
        % next
        \node[draw, circle, minimum size=1cm ] (I1) at (4, 6){$I_{\ell_2
                        m_2\ell' m'}$}; \node[inner sep=0, minimum size=0]() at
                        (5.2, 6) {\tiny $\ \ell_2, m_2 \ $};

        % arrows
        \draw[-] (Y.east) -- (W1.west)
        node[midway,above]{$(\cdot)\times(\cdot)$}; \draw[-stealth] (inv1.south)
        -- (I.north) node[midway,left, anchor=north
        east]{};%{\healpix.\maptoalm$(\cdot)$};
        \draw[-] (Y2.east) -- (W2.west)
        node[midway,above]{$(\cdot)\times(\cdot)$}; \draw[-stealth] (inv2.south)
        -- (I1.north) node[midway,left, anchor=north
        east]{\healpix.\maptoalm$(\cdot) \hspace{10pt}$}; \draw[-] (I.east) --
        (I1.west) node[midway, above] (here) {\tiny
        $(\cdot)(\cdot)^*\times(\cdot)(\cdot)^*$};

        % final node
        \node[draw, diamond, below = of here] (final)
        {$\Theta^{\ell\ell'}_{\ell_1\ell_2}$}; \node[inner sep=0, minimum size=0
        , left of=final]() {\tiny $\ell_1, \ell_2 \ \ \ $};

        \draw[double, -stealth]
        (here.south) -- (final.north) node[midway, right]{\tiny $\sum_{m m'
                    m_1m_2}$};
    \end{tikzpicture}
    \caption{Diagram showing the algorithm to compute the reduced covariance
    coupling kernels using \healpix tools. We use the same notation as in the
    diagram of Fig.~\ref{fig:diag}. The \healpix functions require
    $\mathcal{O}(\nside^3)$, with $\nside$ the chosen resolution of maps $W$ and
    $Y_{\ell m}$. As we choose the resolution \nside to be of the order of the
    multipoles indices $\ell, \ell'$, it is equivalent to say that they require
    $\mathcal{O}((\ell+\ell')^3)$. As operations are done
    $\mathcal{O}(\ell+\ell')$ times, the whole operation of computing
    $\bThe_{\ell\ell'}$ is $\mathcal{O}((\ell+\ell')^4)$ . Finally, it is clear
    in this diagram that the computing time of this kernel is at least doubled
    if one uses multiple masks. One would have different masks as inputs in the
    first line. As a result, one would have to compute the coupling coefficients
    for each of the masks, as shown in Eq.~(\ref{eq:thetamultiplemasks}).}
    \label{fig:diag_thetas}
\end{figure}

Our expression of the covariance matrix approximations in terms of the
normalized coupling kernel $\bThe$ in Eq.~(\ref{eq:NKAThetaform}) gives us a
very efficient tool for examining the validity of each approximation and better
understand their differences. We designed an algorithm to calculate this kernel
exactly similar to the one described in Sect.~\ref{sec:exact} for the exact
calculation of the matrix. We show a diagram of the algorithm in
Fig.~\ref{fig:diag_thetas}.

\begin{figure*}
    \centering
    \includegraphics{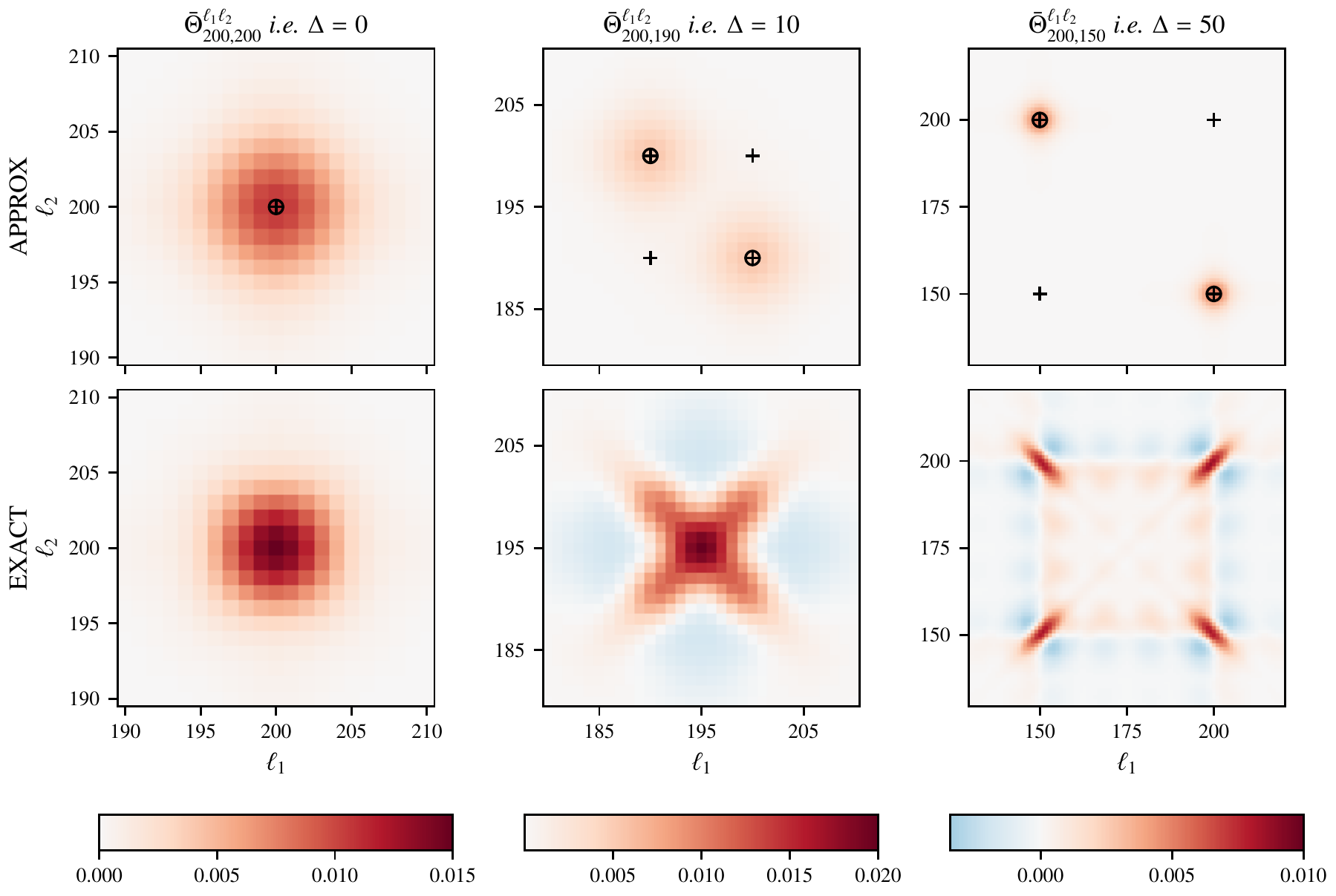}
    \caption{Reduced covariance coupling kernels
    $\smash{\bThe^{\ell_1\ell_2}_{\ell\ell'}}$ for $\ell=200$ and
    $\ell'=\ell+\Delta$, with $\Delta=[0, 10, 50]$ shown in the three different
    columns. All kernels in one column share the same colormap. All of the
    matrices are shown as a function of $\ell_1$ and $\ell_2$ and are centered
    in $\ell_1,\ell_2=(\ell+\ell')/2$. Each of the displayed kernels is properly
    normalized according to Eq.~(\ref{eq:norm_fourcX}). The plots are restricted
    to the elements which have a significant value. The top row shows the
    approximated \INKA kernels and the positions where the delta distributions
    of the \NKA (circles) and \FRI (crosses) approximation peak. The bottom row
    shows the exact kernels. The comparison between the two highlights how much
    of the structure of the exact kernel is missed by the different
    approximations.}
    \label{fig:allfourc}
\end{figure*}

The reduced covariance coupling kernel is then displayed in
Fig.~\ref{fig:allfourc} for the INKA, NKA and FRI approximations compared to the
exact computation for a fiducial multipole $\ell=200$. The kernels are
represented as matrices as a function of $\ell_1$, $\ell_2$, for different fixed
choices of the indices $\ell,\ell'$. Columns show the results for different
choices of $\ell'=\ell-\Delta$, with $\Delta=0$ (i.e. the kernels for the
diagonal terms of the covariance matrix, e.g. $\tilde{\Sigma}_{200,200}$), or
$\Delta=10, 50$ (i.e. the kernels for the off-diagonal terms separated by $10$
or $50$ multipoles, e.g. $\tilde{\Sigma}_{200,190}$). We remind the reader that
the reduced kernel is multiplied to $C_{\ell_1},C_{\ell_2}$ and summed over the
indices $\ell_1, \ell_2$ in Eq.~(\ref{eq:covwithbarX}). Hence,
Fig.~\ref{fig:allfourc} directly shows the weight of the $\ell_1,\ell_2$ power
spectra which contribute to the $\tilde{\Sigma}_{\ell \ell'}$  element of the
covariance matrix.

\begin{figure}
    \centering
    \includegraphics{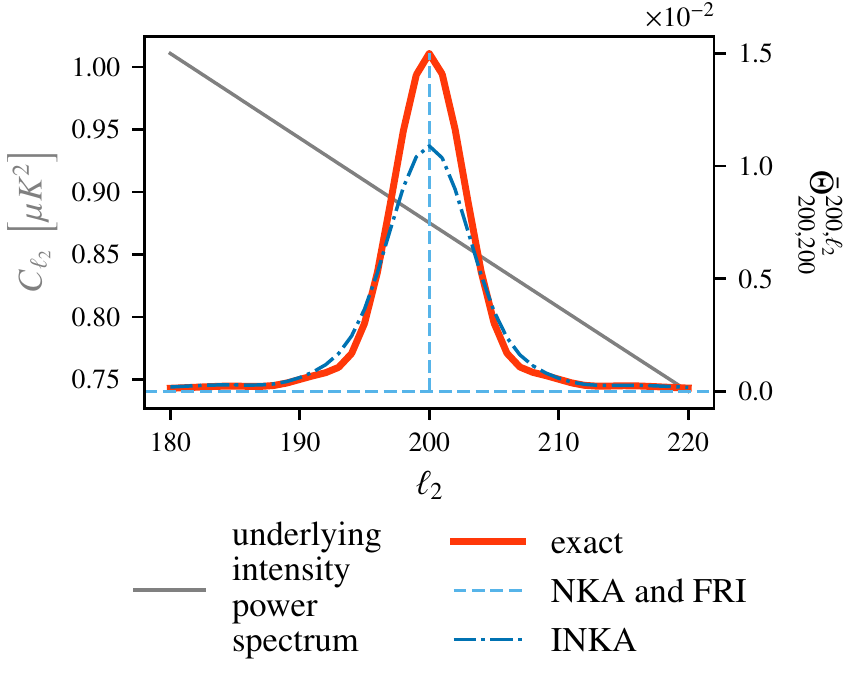}
    \caption{Slices of the exact covariance coupling kernel (solid red) versus
        the approximated \NKA/\FRI (dash light blue) and \INKA (dot-dash dark
        blue) ones (\rhs scale). We show for comparison the CMB intensity power
        spectrum (\lhs scale, solid gray line). The typical width of the exact
        coupling kernel for a small sky footprint is large enough that the power
        spectrum cannot be considered as nearly constant, as required by the
        \NKA/\FRI approximations. The \INKA kernel is lower that the exact one
        at the maximum but have larger tails. This explains why the INKA
        covariance mainly deviates on the peaks or troughs of the power
        spectrum, where it  underestimates or overestimates (respectively) the
        convolution of the kernel with the power spectrum.}
    \label{fig:weightsincoupling}
\end{figure}

We first focus on the kernels for the diagonal of the covariance matrix,
$\Delta=0$, shown on the first column of Fig.~\ref{fig:allfourc}. All kernels
peak at $\ell_1=\ell_2=\ell$, as expected.  However, it is clear from the exact
calculation that the kernel has a significant width compared to the CMB power
spectrum. This is  more clearly shown in Fig.~\ref{fig:weightsincoupling}, where
we plot a slice of the coupling kernels for $\ell_1=200$. One can see that the
width of the kernel cannot be neglected compared to the slope of the CMB
power spectrum. This justifies the \INKA approximation, which replaces the Dirac
$\delta$ functions of \NKA and \FRI by renormalised mode coupling kernels, see
Eq.~(\ref{eq:INKA}). However, as shown in this figure, the \INKA kernel is
slightly larger compared to the exact calculation and of smaller amplitude. This
explains why \INKA underestimates the covariance diagonal in the peaks of the
power spectrum and overestimates it on the troughs, as shown in
Fig.~\ref{fig:approxvsexact_pseudo}, since it averages the underlying power
spectrum in a larger range of multipoles. Conversely, the \NKA/\FRI kernels are
much thinner compared to exact computation, and so they overestimate the
diagonal in the peaks and underestimate it in the troughs of the power spectrum.

Second, we focus on the off-diagonal terms, $\Delta=10,50$, shown in the second
and third column of Fig.~\ref{fig:allfourc}. The difference between the exact
computation and all of the existing approximations is striking and it is clear
that the kernel has more structure than the simple approximated forms. For close
off-diagonal terms such as $\Delta= 10$, the true kernel peaks at its central
index $\ell_1=\ell_2=\bar{\ell}\equiv(\ell+\ell')/2$. For far off-diagonal terms
such as $\Delta=50$, there are four maxima as predicted by the \FRI
approximation, which are partially missed by the \INKA approximation. Moreover,
the true coupling has more dynamics and covers also negative values. This is the
reason why the different approximations, even the \INKA one, fail to correctly
represent the off-diagonal terms of the covariance, as observed in
Fig.~\ref{fig:highmultipolerows_old}.

%%%%%%%%%%%%%%%%%%%%%%

%%%%%%%%%%%%%%%%%%%%%%
\section{A new approximation for the covariance}
\label{sec:newapprox}
\subsection{Improved Approach: the Approximated Covariance Coupling (ACC)}
\begin{figure}
    \centering
    \includegraphics{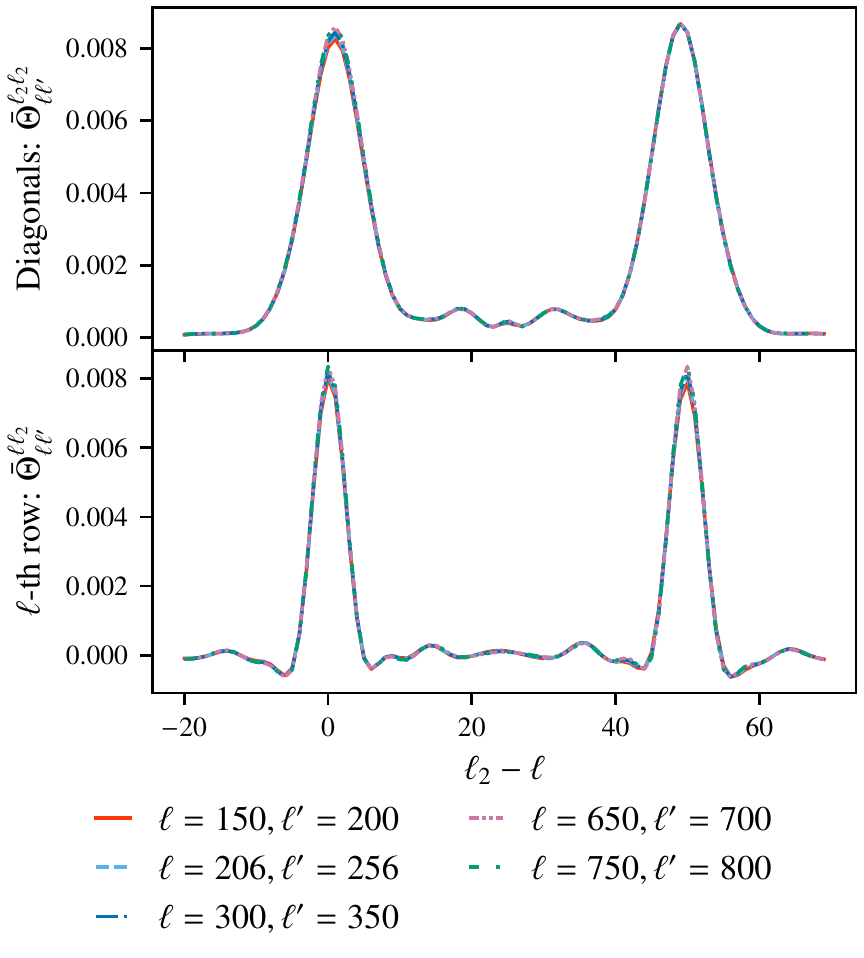}
    \caption{Slices of the reduced covariance coupling kernels
        $\bThe_{\ell,\ell'}^{\ell_1,\ell_2}$ for $\ell \in [150, 206, 300, 650,
        750]$, $\Delta=\ell'-\ell=50$ and $\Delta_1=\ell_1-\ell=0$, as a
        function of $\Delta_2=\ell_2-\ell$. The plots show that for different
        $\ell$ but same $\Delta$, the kernels are very similar, differing only
        at the 5\% level. A similar result can be shown for other values of
        $\Delta$. This leads us to formulate our new approximation, where we
        assume $\bar{\Theta}$ to depend only on the multipole separations
        $\Delta, \Delta_1, \Delta_2$. }
    \label{fig:invariant}
\end{figure}

Our ability to calculate the exact reduced covariance coupling matrix $\bThe$,
described in Section \ref{sec:oldapprox}, allows us to introduce a new
approximation for the computation of the pseudo-power spectrum covariance
matrix. Indeed, we notice that for a fixed $\Delta=\ell'-\ell$, the structure of
$\bThe_{\ell\ell'}$ seems invariant. In other words, the coupling matrices
contributing to the $\tilde\Sigma_{\ell\ell'}$ term of the covariance matrix
only depend on the distance $\Delta$ from the diagonal. This is demonstrated in
Fig.~\ref{fig:invariant}, where we plot diagonal and slices of the exact
calculation of $\bThe_{\ell,\ell+\Delta}$ for $\Delta=50$ and different $\ell$,
for $\ell_1=\ell$. The plot reveals that the kernels are nearly identical when
plotted as a function of $\Delta_2=\ell_2-\ell$. We thus infer that in general
if one writes the $\bThe$ matrices as a function of $\Delta_1=\ell_1-\ell$ and
$\Delta_2 =\ell_2-\ell$, these only depend on $\Delta=\ell'-\ell$, for any
$\ell$ and $\ell'$. The difference between kernels computed at different
$\ell$'s for same $\Delta$ is small, at the 5\% percent level. We can thus
assume that for any choice of multipole $\ell,\lambda$:
\begin{align}
    \forall (\ell, \lambda),
    \bThe_{\ell(\ell+\Delta)}^{(\ell+{\Delta_1})(\ell+{\Delta_2})}
    \approx
    \bThe_{\lambda(\lambda+\Delta)}^{(\lambda+{\Delta_1})(\lambda+{\Delta_2})}.
\end{align}
An analytical justification of this approximation is provided in
App.~\ref{sec:appacc} using the asymptotic expansion of the Wigner-3j symbols
when $\ell$ is large.

This suggests a new approximation where the coupling kernel just has to be
computed at a given fiducial $\ell$ for all relevant values of $\Delta,\Delta_1$
and $\Delta_2$.  We will denote this new \emph{approximated covariance coupling}
method as \ACC. More precisely, the \ACC kernel is given by
\begin{equation}
    \label{eq:approxTheta}
    \bThe_{\ell\ell'}^{\ell_1\ell_2}
    \approx \bThe_{\ell(\ell + \Delta);\ACC}^{(\ell + \Delta_1)(\ell + \Delta_2)}
    \equiv \bThe_{\ell^* (\ell^*+\Delta)}^{(\ell^* + \Delta_1)(\ell^* + \Delta_2)},
\end{equation}
where we perform the exact and costly computation only for the
$\bThe_{\ell^*(\ell^*+\Delta)}^{(\ell^* + \Delta_1)(\ell^* + \Delta_2)}$. We are
free to choose the reference $\ell^*$ multipole. Since there are no significant
long range correlations in our case, we can pick a low\footnote{in the limit
where the asymptotic justification of App.~\ref{sec:appacc} remains valid.}
$\ell^*$ and use a low $\nside$ map resolution. We have to make sure, however,
that $\ell^*$ is larger than $\lcut$, the low-$\ell$ cutoff that was introduced
to avoid issues with large scale leakages. Close to $\lcut$, the exact
computation can be used. Note that with a small mask, large scale modes are
difficult to measure and usually excluded from the cosmological analysis. We can
also restrict the range of $\Delta$ to the number of off-diagonal terms of
interest in the covariance matrix. The correlation falls quickly (see
Figs.~\ref{fig:rowsimsvsexact} and~\ref{fig:2dcov}) and, in practice, we can
restrict it to $|\Delta|<d_\mx$, with $d_\mx$ being of the order of a few times
the correlation length. Similarly, the kernels fall quickly in
$\Delta_1,\Delta_2$,  so we can also restrict ourselves to a small region of a
similar order and, in the case of the single mask analysis, use the symmetry
around $\Delta \leftrightarrow -\Delta$ to reduce the cost of the computation.

While we only presented temperature coupling kernels in Fig.~\ref{fig:allfourc},
the situation is identical in polarization, and a similar approximation can be
built, see App.~\ref{sec:appacc}. We used this approximation with $\ell^*=300,\
\nside=512$ and $d_\mx=100$ to compute the \ACC results in
Fig.~\ref{fig:approxvsexact_pseudo}.

\subsection{Accuracy and scaling}
\label{subsec:newaccuracy}
\begin{figure}
    \centering
    \includegraphics{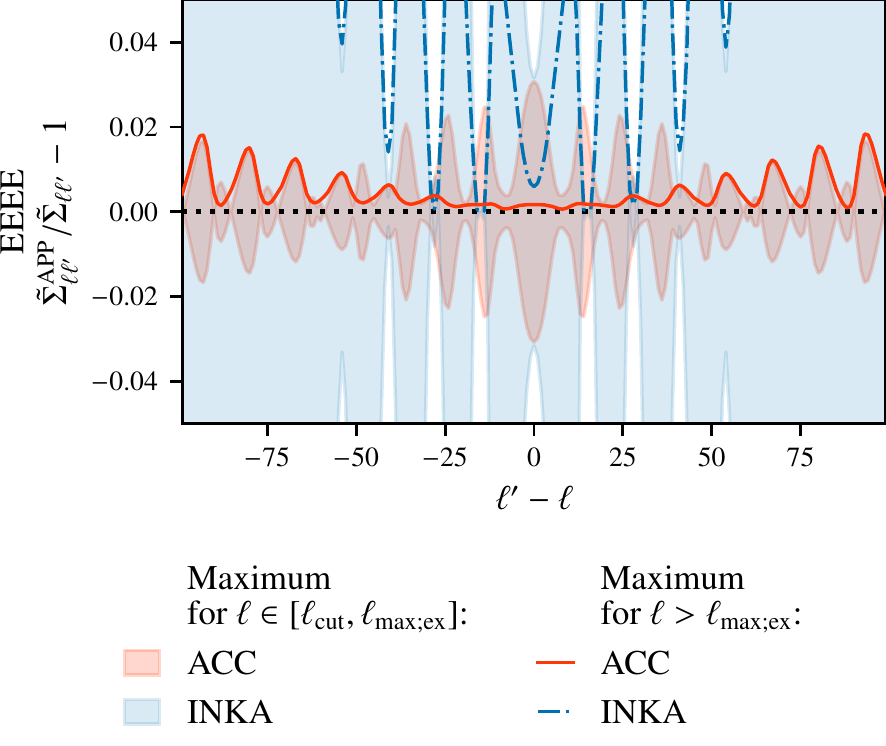}
    \caption{Zoom of Fig.~\ref{fig:highmultipolerows_old}. We focus on the
        relative differences of the \ACC and \INKA covariance matrices with
        respect to the exact covariance for \EEEE. The deviations found at
        $\ell<1000$ (shaded regions) are similar to the ones found at the higher
        multipoles (lines), showing that the approximations work at the same
        level of accuracy in the two cases.}
    \label{fig:highmultipolerows}
\end{figure}

\begin{table*}
    \centering
    \begin{tabular}{ccccc}
        \hline
        Method                                                           &
        Mathematical definition                                          &

        $\bThe_{\ell\ell'}^{\ell_1\ell_2}$                               &

        Precision                                                        &

        Complexity \\
        \hline\hline
        Exact (this work)                                                &
        \ref{eq:covwithbarX}                                             &
        $\bThe_{\ell\ell'}^{\ell_1\ell_2}$ computed $\forall \ell, \ell',
        \ell_1, \ell_2$                                                  & N/A &
        $\mathcal{O}(\lmax^5)$ \ \ (using \healpix pixelation) \\
        \hline\hline
        \NKA                                                             &
        \ref{eq:NKA}
                                                                         &
        $(\delta_{\ell\ell_1}\delta_{\ell'\ell_2} +
        \delta_{\ell'\ell_1}\delta_{\ell\ell_2})/2$                      & 4\% &
        $\mathcal{O}(1)$ \\
        \hline
        \FRI                                                             &
        \ref{eq:FRI}
                                                                         &
        $(\delta_{\ell\ell_1} + \delta_{\ell'\ell_1}) (\delta_{\ell\ell_2} +
        \delta_{\ell'\ell_2})/4$                                         & 4\% &
        $\mathcal{O}(1)$ \\
        \hline
        \INKA                                                            &
        \ref{eq:INKA}
                                                                         &
        $({}_0\bar{M}_{\ell\ell_1}{}_0\bar{M}_{\ell'\ell_2} +
        {}_0\bar{M}_{\ell'\ell_1}{}_0\bar{M}_{\ell\ell_2})/2$            & 4\% &

        $\mathcal{O}(\lmax^3)$ or $\mathcal{O}(\lmax^2)$ with
        \cite{Louis2020}
        \\
        \hline
        \ACC (this work)                                                 &
        \ref{eq:approxTheta}                                             &
        $\bThe_{\ell\ell'}$ invariant for $\Delta\equiv|\ell-\ell'|=$cst & 1\% &
        $\mathcal{O}(d_\mx \nside^4)$ \\
        \hline
    \end{tabular}
    \caption{Summary of computation methods to obtain the pseudo-power spectrum
    covariance matrix. {First column:} Name of approximation. {Second column:}
    Equation to which they are referred. {Third column:} Expression of $\bThe$
    in this approximation. {Fourth column:} Precision determined by the maximum
    values of the relative difference of the \EEEE binned covariance on diagonal
    for multipoles $\lcut\leq\ell\leq \lexact$ in
    Fig.~\ref{fig:approxvsexact_pseudo}. For larger multipoles, the
    approximation are expected to be in this range of precision as shown in
    Fig.~\ref{fig:highmultipolerows}. {Fifth column:} Summary of computing
    resources needed to obtain $\bThe$ in each approximation. Let us here
    specify that for \INKA, the kernel $\bar{M}$ is often already known, thus
    the practical complexity is $\mathcal{O}(1)$. $\lmax$ is the multipole range
    of the covariance, $d_\mx$ is the number of diagonal computed in the \ACC
    approximation, \nside is the resolution chosen to compute the covariance
    coupling kernels in the \ACC approximation (closest to $\lcut$ is
    sufficient).}
    \label{tab:summary}
\end{table*}
We validate the accuracy of our \ACC approximation and compare it to the other
approximations in Fig.~\ref{fig:approxvsexact_pseudo}. Our new approximation
succeeds at estimating the covariance within 1\% error for all multipoles larger
than $\lcut$, in intensity and polarization, which is a factor of $\sim 4$
improvement over previous approximations. This is also shown in
Fig.~\ref{fig:highmultipolerows_old} and in Fig.~\ref{fig:highmultipolerows},
which is just the same as Fig.~\ref{fig:highmultipolerows_old} but focused on
the \EEEE \ACC residuals with respect to \INKA. It is clear from this figure
that the \ACC approximation estimates much better both the diagonal and
off-diagonal terms of the covariance matrix.

Figure~\ref{fig:diag_thetas} shows the computations needed to obtain the
covariance coupling kernels. Following the same argumentation as for the exact
computation of the Sect.~\ref{sec:exact}, we can show that the computation of a
single kernel $\bThe$ scales as $\mathcal{O}((\ell+\ell')^4)$. As a result,
since we need to compute one for each diagonal index $\Delta \in [0, d_\mx]$,
the final \ACC approximation scales as $\mathcal{O}(\nside^{4} d_\mx)$, where
$\nside$ is the map resolution chosen to compute the kernels. The computing
resources needed to obtain $\bThe$ for all approximations are summarized in
Table~\ref{tab:summary}. They add up to the resources needed to compute the
symmetric coupling kernel $\Xi$ in Eq.~\ref{eq:covwithbarX}. In practice, the
kernel $\bar{M}$ needed to build \INKA is often already known for the sky
analysis, as it has the same structure as $\Xi$, so the effective complexity for
this approximation is $\mathcal{O}(1)$.

\subsection{Point source mask}

We did not include a point source mask in the survey footprint. Point source
masks significantly complicate the problem as the power spectrum of the mask
will have power at large multipoles, hence it will extend the correlation
length. This has been an issue for all analyses thus far. Apodizing the point
source masks helps to alleviate the problem but at the price of discarding a
significant area of the usable sky. Even in the case of a large survey
footprint, such as \planck, the point source masks have been shown to break the
\NKA approximation. In this case the issue was mitigated using a
simulation-based correction. Concerning \FRI, \INKA and \ACC, preliminary work
that we have performed also suggest that they fail when including sources. We
expect these approximations to perform poorly  since the improvements over \NKA
are focused on the central shape of the reduced covariance coupling matrix
$\bThe$, while the sources tend to affect the far off-diagonal terms, increasing
the correlation between distant multipoles. Looking particularly at the \ACC
approximation, the larger correlations at distant multipoles will break the
asymptotic behavior of the Wigner-3j symbols shown in App.~\ref{sec:appacc}. We
can expect that this will reduce the validity of the approximated invariance by
translation along the covariance diagonal, which is at the core of the \ACC
approximation. More work is required to assess the accuracy of \ACC and other
approximations in this case. However, one can adopt different  approaches to
mitigate the effect of a point source mask. For example, one could find
analytical solutions (\citet{Gratton}, in prep.), inpaint the maps
\citep{Benoit-Levy2013}, or use a Monte-Carlo correction such as the one used in
\planck \citep{PlanckCollaboration2015}.

%%%%%%%%%%%%%%%%%%%%%%

%%%%%%%%%%%%%%%%%%%%%%
\section{Covariance of the \polspice estimator}
\label{sec:polspice}
The pseudo-power spectrum is a biased estimator of the true underlying spectrum
of the masked \cmb maps. To recover an unbiased estimator, one can apply the
\master\ \citep{Hietal02} formalism, which inverts the mode-coupling matrix and
applies it to the biased estimator. Similarly, one can use the
\polspice\citep{Szapudi2001, Chonetal04} algorithm, which corrects  the two
point correlation function for the effect of the mask in real space and then
converts the result back into harmonic space. However, when the sky footprint is
small, and large angular scales are not observed, the unbinned mode-coupling
matrix becomes non-invertible. Analogously, the \polspice\ conversion of the
two-point correlation function into a power spectrum cannot be performed. In the
first case, the mode-coupling matrix must be binned to allow the inversion. In
the second case, one must apodize (i.e. gradually cut) the large angular scales
of the two-point correlation function before calculating the corresponding power
spectrum. This introduces a small bias in the final estimator which cannot be
corrected for.

In this section, we will explain in detail how to calculate the covariance matrix for the
\polspice estimator starting from a pseudo-power spectrum covariance matrix,
which we will produce through our \ACC approximation. We will show how to
include the effect of the correction of the mask, as well as the small bias
introduced by the \polspice apodization of the two-point correlation function.
In particular, we will show that this apodization can be expressed in harmonic
space, allowing us to relate the \polspice spectrum covariance matrix to the the
pseudo-spectrum one with a convolution.

%%%
\subsection{\master equation}
The pseudo-power spectrum is related to the true one through the well-known
\master equation introduced in \cite{Hietal02}
\begin{equation}
    \label{eq:masterT}
    \langle \tC_{\ell}^{\TT} \rangle =
    \sum_{\ell'} {}_0M_{\ell\ell'} C_{\ell'}^{\TT},
\end{equation}
with similar equations for polarization, see App.~\ref{sec:appmasterpol}. This
bias comes from the missing information due to the masked sky. Given the
weighted mask $W(\hat{n})$, one can compute ${}_0M$ using
Eq.~(\ref{eq:defineM}). Provided that ${}_0M$ is invertible, an unbiased
estimator can be constructed. These relations can be expressed in real space
using the two-point correlation functions $\xi$, which for a statistically
isotropic sky depend only on the relative angle between two directions
\begin{equation}
    \langle T(\hat{n}_1)T(\hat{n}_2)\rangle =
    \xi(\arccos(\hat{n}_1\cdot\hat{n}_2)).
\end{equation}
They can be related to the power spectrum $C_\ell$ using a Legendre series, with
\begin{align}
    \label{eq:spectocorr}
    \xi(\theta)
           & = \sum_\ell \frac{2\ell+1}{4\pi}
    C_\ell P_\ell(\cos\theta),                                                 \\
    C_\ell & = 2\pi \int_0^\pi \ud\cos\theta  \xi(\theta) P_\ell(\cos \theta).
\end{align}
If we define in the same manner the correlation function $\tilde{\xi}$ of the
masked sky, associated with the pseudo-spectrum $\tC_\ell$, we obtain from
Eq.~(\ref{eq:masterT}), by applying the decomposition in a Legendre series, the
following relation
\begin{equation}
    \label{eq:realmasterT}
    \langle \tilde{\xi}(\theta)\rangle  = w(\theta) \xi(\theta), \
    \forall \theta \in [0, \pi],
\end{equation}
where $w(\theta)$ is the mask angular correlation function (more details can be
found in App.~\ref{sec:applegendre}). From this relation, we can establish that
the \master mode-coupling matrix ${}_0M$ is invertible only if the correlation
function of the mask $w(\theta)$ is non-zero for all $\theta \in [0, \pi]$,
which implies that the survey area explores all angular separations on the sky.
While this is valid for almost full sky analyses such as the \planck one, it does
not hold for experiments observing small patches, such as \spt, where angular
scales larger than $\theta \sim 30\deg$ are unexplored. As a result, ${}_0M$ is
not invertible. Binning allows the regularization of the \master matrix and thus to
build a nearly-unbiased estimator of the bandpowers. This approach is described
in \cite{Hietal02} and is adopted in
\texttt{NaMaster}\footnote{\url{https://github.com/LSSTDESC/NaMaster}}
\citep{Alonso2019}. Similarly, we show in the next section how the unobserved
large angular scales are handled in the \polspice estimator.

%%%%
\subsection{Regularizing with PolSpice}
\subsubsection{Temperature}
The pseudo-power spectrum estimator can be regularized in real space following
the \polspice approach in \cite{Szapudi2001}. The pseudo-correlation function
$\tilde{\xi}$ is smoothed with a scalar apodizing function $\fapo(\theta)$,
which cuts out large $\theta$, and then corrected for the bias coming from the
weighted mask described in Eq.~(\ref{eq:realmasterT}). The scalar apodizing
function goes smoothly from $\fapo(0)=1$ to $\fapo(\theta_{\mx})=0$ to avoid
Fourier ringing. $\theta_\mx$ should be chosen as the maximal angular size of
the weighted mask. A new correlation function estimator $\hat{\xi}(\theta)$ is
defined as
\begin{align}
    \label{eq:polspicecorr}
    \hat{\xi}(\theta) & \equiv g(\theta) \tilde{\xi}(\theta),
\end{align}
with
\begin{equation}
    g(\theta) = \left\{
    \begin{aligned}
         & \fapo(\theta)/w(\theta) \
         & \forall \theta \in \left[0, \theta_{\mx}\right) \ , \\
         & 0  \
         & \forall \theta \in \left[\theta_{\mx}, \pi\right].
    \end{aligned}\right.
\end{equation}
The function $g$ is well defined and smooth for all angles thanks to the
apodization $\fapo$. As a consequence of Eq.~(\ref{eq:realmasterT}) and
Eq.~(\ref{eq:polspicecorr}), the \polspice estimator of the correlation function
can be related on average to the true underlying correlation function with
\begin{align}
    \langle \hat{\xi}(\theta) \rangle & = \fapo(\theta) \xi(\theta) \ \
    \forall \theta \in \left[0, \pi\right].
\end{align}
Going back to harmonic space using a Legendre transform, this operation can be
expressed as:
\begin{align}
    \langle\hC_\ell \rangle & = \sum_{\ell'} {}_0K_{\ell\ell'} C_{\ell'}. \\
    {}_0K_{\ell\ell'}       & =(2\ell'+1)\Xi^{00}_{\ell\ell'}[\fapo]
\end{align}
The \polspice kernel ${}_0K$ is obtained from the scalar apodizing function
$\fapo$ with an extended definition of the operators $\Xi$ (see
App.~\ref{sec:applegendre} and Eq.~(\ref{eq:definexioncorrfunc}) for more
details). The operator acts on the Legendre transform of $\fapo$.

The advantage of \polspice, which performs the regularization in real space
rather than in harmonic space, is that it replaces an $\ell$-space convolution
by an integration and a multiplication, which are faster and numerically more
stable, producing an estimator for all multipoles $\ell$. We will denote this
estimator with a hat, for instance $\hC_\ell^{\rm \textsc{XY}}$. Note that this
regularization (which is only required for small sky patches) introduces a bias
in the \polspice estimator that cannot be corrected for. The bias is small,
since ${}_0K$ is properly normalized, \ie $\sum_{\ell\ell'}
{}_0K_{\ell\ell'}=1$. Furthermore, the regularization increases the
correlations between unbinned modes. The \polspice kernels behave as window
functions, mixing multipoles of the pseudo-power spectrum. The lack of
information at large scale induces the inability to distinguish multipoles which
are close to each other. For this reason, for cosmological analyses, the
spectrum estimator is binned in ranges larger than the typical correlation
between multipoles.

%%%
\subsubsection{Polarization}
\polspice allows one to correct for the bias introduced by the cut sky in the
same manner for the polarized spectra. It also allows one to decouple the \EE
and \BB estimator, see \cite{Challinor2004a} or the appendix of this work for
details, App.~\ref{sec:applegendre}. Similarly to the intensity case, we can
express the effect of the \polspice real space regularization in spherical
harmonics by defining the polarized \polspice kernels: ${}_{\pm2}K_{\ell\ell'} =
(2\ell'+1)\Xi^{2\pm2}_{\ell\ell'}[\fapo]$. The \polspice estimator follows for
\textsc{X} $\in \left[\E, \B\right]$
\begin{align}
    \langle\hC_\ell^{\rm \textsc{XX}} \rangle =
    \sum_{\ell '}{}_{-2}K_{\ell\ell'} C_{\ell'}^{\rm \textsc{XX}}.
\end{align}
Concerning the temperature$\times$polarization case, we can show that
\begin{align}
    \langle\hC_\ell^{\rm \textsc{TE}} \rangle & =
    \sum_{\ell '}{}_{\times}K_{\ell\ell'} C_{\ell'}^{\rm \textsc{TE}},                   \\
    \text{ with } {}_{\times}K_{\ell\ell'}    & = (2\ell'+1)\Xi_{\ell\ell'}^{20}[\fapo].
\end{align}

%%%
\subsection{Relating \polspice and \master in harmonic space}
We can translate the relations Eq.~(\ref{eq:polspicecorr}) into harmonic space
in temperature and polarization to obtain the \polspice estimator as a harmonic
convolution of the pseudo-power spectrum estimator
\begin{align}
    \hC^{\TT}_\ell                 & = \sum_{\ell '} {}_0G_{\ell\ell'} \tC^{\TT}_{\ell'},
    \label{eq:clspiceGltt}                                                                \\
    \hC^{\EE}_\ell -\hC^{\BB}_\ell & =
    \sum_{\ell '}
    {}_{-2}G_{\ell\ell'} \left(\tC^{\EE}_{\ell'} - \tC^{\BB}_{\ell'}\right).
    \label{eq:clspiceGlpolminus}                                                          \\
    \hC^{\EE}_\ell +\hC^{\BB}_\ell & =
    \sum_{\ell '}
    {}_{\mathrm{dec}}G_{\ell\ell'}
    \left(\tC^{\EE}_{\ell'} + \tC^{\BB}_{\ell'}\right),
    \label{eq:clspiceGlpolplus}                                                           \\
    \label{eq:clspiceGlte}
    \hC^{\TE}_\ell                 & =
    \sum_{\ell '} {}_{\times}G_{\ell\ell'} \tC^{\TE}_{\ell'}.
\end{align}

The $G$ kernels are constructed in the same manner as the \polspice kernels,
with the operator $\Xi$ acting on the function $g = \fapo/w$ accordingly to
Eq.~(\ref{eq:definexioncorrfunc}) (or equivalently on the associated power
spectrum of $g$ via Legendre transform). They are given by
\begin{align}
    {}_0 G_{\ell\ell'}             & = (2\ell'+1) \Xi^{00}_{\ell\ell'}[g],    \\
    {}_{-2} G_{\ell\ell'}          & = (2\ell'+1) \Xi^{2-2}_{\ell \ell'} [g], \\
    {}_{\times}G_{\ell\ell'}       & = (2\ell'+1) \Xi^{20}_{\ell\ell'}[g],    \\
    {}_{\mathrm{dec}}G_{\ell\ell'} & = \frac{2\ell'+1}{2}
    \int_{-1}^1 g(\theta) d^\ell_{22}(\theta)
    d^{\ell'}_{2-2}(\theta) \ud \cos(\theta). \label{eq:gdec}
\end{align}

The first three equations above reduce to the inverse of the master kernels when
the \polspice apodization function is set to 1, \ie no apodization. The last
kernel, referred to as \emph{dec} is the kernel that allows the decoupling of the \EE
and \BB spectra. The appendix~\ref{sec:applegendre} gives more details on this
point. ${}_{\mathrm{dec}}G$ is associated with integral relations in real space,
thus its harmonic expression is not straightforward. This expression requires
more numerical resources to be computed since there is no closed relations for
the Wigner d-matrix with different multipole indices. It can be obtained with
\polspice for all $\ell$, setting as input $\tC_{\ell'}^{\EE} =
\tC_{\ell'}^{\BB} = \delta_{\ell\ell'}$.

%%%
\subsection{Covariance of \polspice estimator}
\begin{figure}
    \centering
    \includegraphics{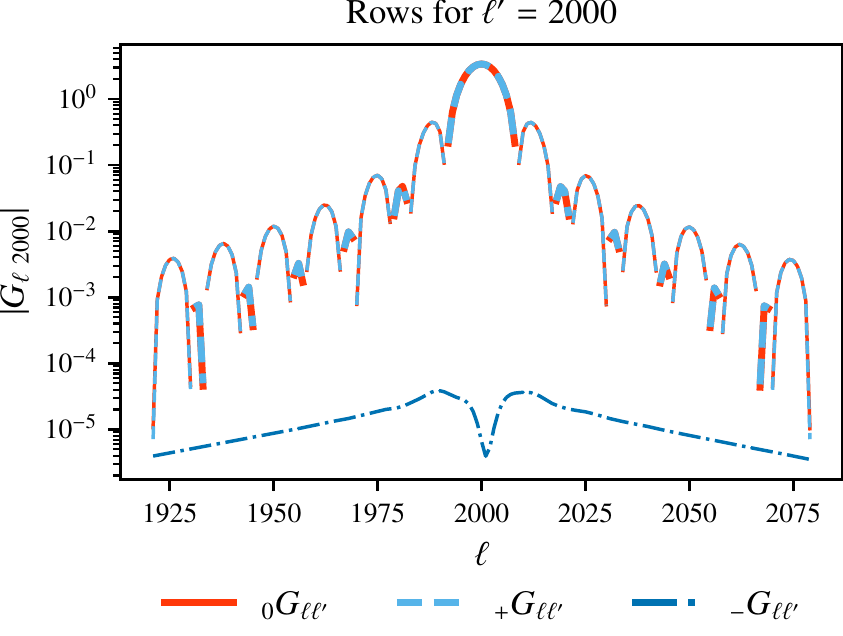}
    \caption{Amplitude of the \polspice convolution kernels $G$ for the \spt
        footprint. The negative terms are plotted with thinner lines. The
        kernels behave as window functions, mixing multipoles of the
        pseudo-power spectrum into the new \polspice spectrum. They correct for
        the bias due to the mask, but introduce a small bias due to the lack of
        information at large scales. The temperature kernel ${}_0G$ and the
        polarization ${}_{+}G$ one are almost identical. The leakage kernel
        ${}_{-}G$ (all negative), which accounts for the mixing of the $E$ and
        $B$ polarization pseudo-spectra in the \polspice spectrum, is orders of
        magnitudes smaller than the other two. Hence the \BB covariance terms do
        not affect the \EEEE covariance. On the other hand, the \EE terms affect
        the \BBBB covariance, as the \EE spectrum is a few orders of magnitude
        larger than \BB.}
    \label{fig:gkernels}
\end{figure}

Given the previous relations in Eq.~(\ref{eq:clspiceGltt}),
(\ref{eq:clspiceGlpolminus}), (\ref{eq:clspiceGlpolplus}) and
(\ref{eq:clspiceGlte}), the covariance of the \polspice estimator can be written
as a convolution of the covariance of the pseudo-power spectrum, with
\begin{align}
    \hSig^\TTTT_{\ell\ell'} & \equiv \cov(\hC^\TT_\ell,\hC^\TT_{\ell'})
    \nonumber                                                           \\
                            & =
    \sum_{LL'} {}_0G_{lL} \ \tSig^\TTTT_{LL'} \ {}_0G_{l'L'} .
    \label{eq:polspicecov}
\end{align}
For polarization, there is mixing between the \EE and \BB components in the
covariance. Let us write down the polarized \EEEE \polspice covariance, after
defining ${}_{\pm}G \equiv \half \left( {}_{\mathrm{dec}}G \pm {}_{-2}G
\right)$, as
\begin{equation}
    \begin{split}
        \label{eq:polspicepolcov}
        \hSig^{\EEEE} =
        &{}_{+}G\tSig^\EEEE{}_{+}G^\top +
        {}_{-}G\tSig^\BBEE{}_{+}G^\top \\
        & + {}_{+}G\tSig^\EEBB{}_{-}G^\top +
        {}_{-}G\tSig^\BBBB{}_{-}G^\top.
    \end{split}
\end{equation}
The polarized \polspice covariance is built on the polarized pseudo-covariance,
mixing the components \EE and \BB, thanks to the kernel ${}_{\pm}G$, see
Fig.~\ref{fig:gkernels}. This figure displays a row of the kernels computed on
the mask \sptyeete used in our analysis. It shows the window functions that are
applied to the pseudo-power spectra to produce the \polspice spectra. The
\polspice apodizing function of the correlation function we used is:
\begin{equation}
    \fapo(\theta) = \left\{
    \begin{aligned}
         & \half \left(1 + \cos \frac{\pi\theta}{\theta_\mx}\right) & \forall \theta < \theta_\mx, \\
         & 0                                                        & \text{ otherwise}.
    \end{aligned}\right.
\end{equation}
Here we have , $\theta_\mx = \pi/6$. Without apodization but with partial sky,
as for \planck, the decoupling kernel ${}_\mathrm{dec}G$ is not null, still
resulting in a non-zero ${}_{-}G$ kernel. However, it will be orders of
magnitude smaller than ${}_{+}G$, hence one can compute \EEEE ignoring the
leakage from covariance terms that include \BB.

\subsection{Accuracy of the covariance approximations for the \polspice estimator}
\begin{figure*}
    \centering
    \includegraphics{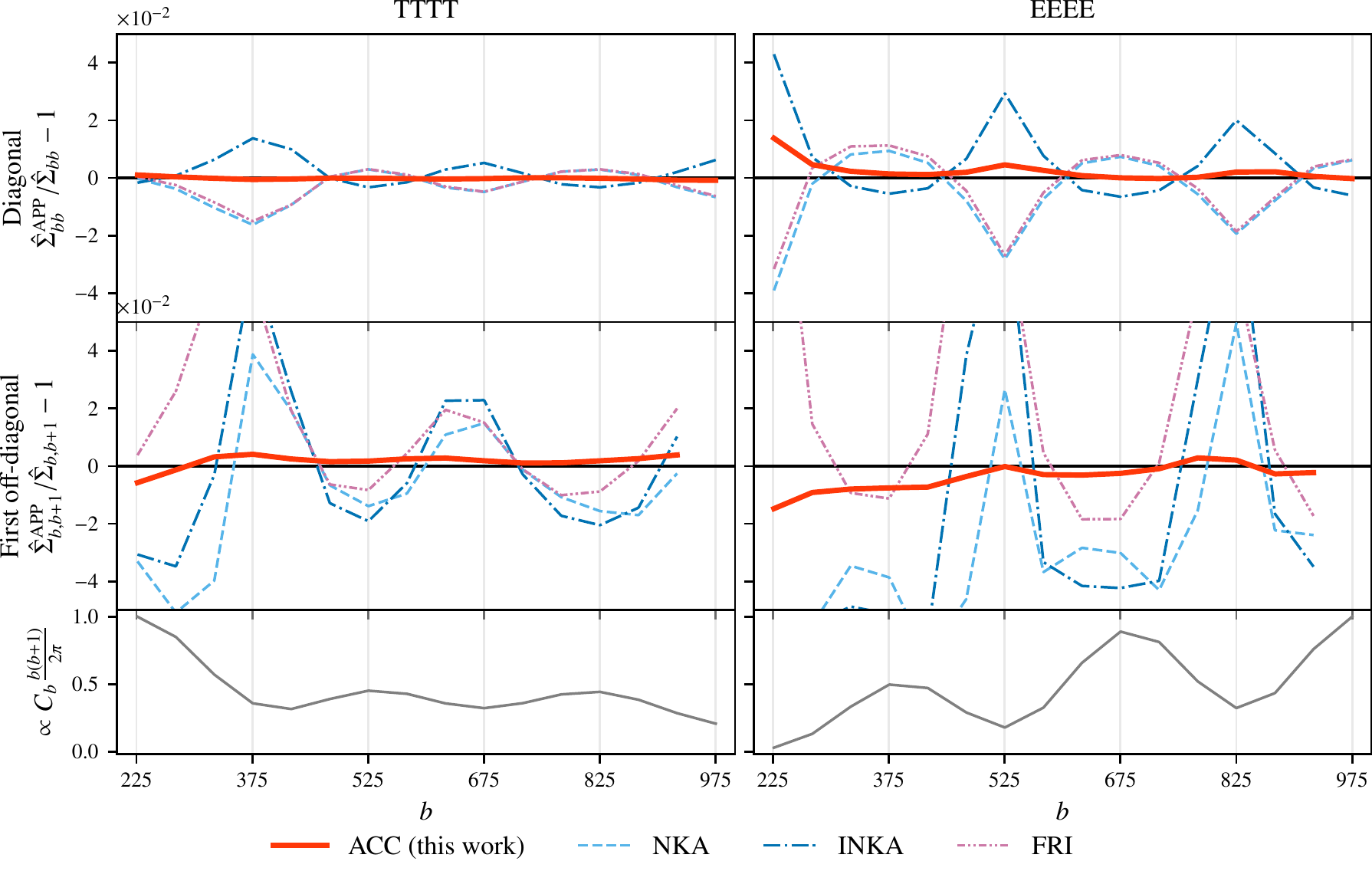}
    \caption{Relative differences of binned \polspice covariance matrices
        calculated using approximations of the pseudo-spectrum covariance with
        respect to the exact computation:
        $\hat{\Sigma}^\mathrm{APP}_{bb'}/\hat{\Sigma}_{bb'} - 1$, for \TTTT
        (\lhs) and \EEEE (\rhs), with binning $\Delta \ell = 50$ . On the first
        row we plot the relative differences for the diagonal, \ie $b=b'$, while
        on the second row we plot the differences for the first off-diagonal,
        \ie $b'=b+1$. Similarly to the case of pseudo-covariances in
        Fig.~\ref{fig:approxvsexact_pseudo}, we find acceptable accuracy for the
        \NKA ({dash light blue}), \INKA ({dash-dot dark blue}) and \FRI
        ({dash-double-dot purple}) approximations, while our \ACC approximation
        ({solid red}) improves over all of the others. The \polspice covariance
        matrices have been calculated using Eqs.~(\ref{eq:polspicecov}) and
        (\ref{eq:polspicepolcov}). The last row displays the corresponding
        binned underlying renormalised spectrum \TT or \EE, to highlight the
        fact that the differences in the covariances are on the peaks and in the
        troughs of the spectra, \ie where the spectral variations are
        maximal.}
    \label{fig:approxvsexact_spice}
\end{figure*}

We can build estimates of the \polspice spectrum covariance convolving the
pseudo-spectrum covariance with the appropriate kernels following
Eqs.~(\ref{eq:polspicecov}) and (\ref{eq:polspicepolcov}). To calculate the
pseudo-spectrum covariance, we can use the \NKA, \INKA, \FRI and \ACC
approximations or the exact computation. Fig.~\ref{fig:approxvsexact_spice}
shows the accuracy of the binned \polspice covariance calculated with the
approximations compared to the exact calculation. The results are similar to the
ones we found for the accuracy of the pseudo-spectrum covariances shown in
Fig.~\ref{fig:approxvsexact_pseudo}. The \NKA, \INKA, \FRI approaches provide  a
good estimate of the \polspice covariance. However, the \ACC approach improves
dramatically over  the existing approximations. This shows that our results for
the accuracy of the pseudo-covariance holds also for the \polspice one.

%%%%%%%%%%%%%%%%%%%%%%

%%%%%%%%%%%%%%%%%%%%%%
\section{Summary and conclusions}
%%%%%%%%%%%%%%%%%%%%%%
\label{sec:conclusions}

One of the key ingredients of cosmological analyses based on power spectra are
covariance matrices. Accurate covariance matrices ensure precise error bars and
an unbiased estimation of cosmological parameters. The analytical estimation of
these matrices can be difficult in the case of small observed sky fractions,
since existing approximations might fail. We have considered the
specific example of estimating accurate analytical covariance matrices for the
\spt \cmb experiment, whose survey covers around $4\%$, without masking the
contribution of point sources. We considered both the cases of estimating the
matrix for pseudo-power spectrum and for the  \polspice power spectrum
estimator.

First, in Sec.~\ref{sec:exact}, we implemented for the first time an expensive
exact calculation of the covariance of the pseudo-power spectrum in intensity
and polarization. We used a map-based algorithm that is accelerated thanks to
the \healpix pixelation tools. We were thus able to compute exactly the entire
covariance matrix up to $\lexact = 1000$. We also obtained a selection of rows
of the covariance of particular interest up to $\ell =2000$.

Thanks to this result, we were able to estimate precisely the accuracy of the
existing approximations in Sec.~\ref{sec:oldapprox} by comparing them to the
binned exact covariances of the pseudo-power spectra measured on the \sptyeete
patch. The approximations were found to be precise to the 5\% level.

Then, using the code we developed for an exact computation of the covariance
matrix, we estimated the covariance coupling kernel \bThe, which determines how
the \cmb power spectrum couples into the covariance matrix. We were able to
understand why the existing approximations in the literature fail to achieve a
precision better than 5\%. We then proposed in Sect.~\ref{sec:newapprox} a new
approximation, the {\it Approximated Covariance Coupling}  (\ACC), which is more
computationally expensive than the existing approximations, but allows one to
have a more precise estimation of the covariance matrix at the 1\% percent
level.

Finally, in Sec.~\ref{sec:polspice}, we showed that we were able to build the
covariance of the \polspice power spectrum in both temperature and polarization
using a harmonic correction. This computation is exact and based on the
\polspice algorithm real space corrections which we translated into harmonic
space. Thanks to this correction, we produced estimates of the \polspice
covariance matrix based on the previous approximations of the pseudo-power
spectrum covariance. The accuracy of the resulting \polspice covariance
approximations is the same as the pseudo-power spectrum case.

While  this paper considered the particular example of the \spt experiment, the
results can be extended to non-\cmb power spectrum analysis such as the one of
weak shear or photometric catalogs. Also, we would like to stress that the
accuracy of any of the approximations presented in this paper (existing or new
one) is reduced once a point source mask is included in the sky footprint.
Nevertheless, the exact computation of the covariance matrices still holds in
this particular case. While previous experiments have included the effect of
point source masks through the use of simulations (see e.g.
\cite{PlanckCollaboration2015}) or by inpainting the holes with constrained
realizations (see e.g. \cite{Benoit-Levy2013}), additional work is required to find
an analytical calculation of this contribution.

\begin{acknowledgements}
We are grateful to the SPT collaboration for discussions and suggestions. We
thank Lennart Balkenhol for discussions and insightful comments on the
manuscript. This work has received funding from the French Centre National
d’Etudes Spatiales (CNES). This project has received funding from the European
Research Council (ERC) under the European Union’s Horizon 2020 research and
innovation programme (grant agreement No 101001897).  
\end{acknowledgements}

\bibliography{camphuis_1}{}

\def\eprinttmppp@#1arXiv:@{#1}
\providecommand{\arxivlink[1]}{\href{http://arxiv.org/abs/#1}{arXiv:#1}}
\def\eprinttmp@#1arXiv:#2 [#3]#4@{\ifthenelse{\equal{#3}{x}}{\ifthenelse{
\equal{#1}{}}{\arxivlink{\eprinttmppp@#2@}}{\arxivlink{#1}}}{\arxivlink{#2}
  [#3]}}
\providecommand{\eprintlink}[1]{\eprinttmp@#1arXiv: [x]@}
\providecommand{\eprint}[1]{\eprintlink{#1}}
\providecommand{\adsurl}[1]{\href{#1}{ADS}}
\begin{thebibliography}{26}
\expandafter\ifx\csname natexlab\endcsname\relax\def\natexlab#1{#1}\fi

\bibitem[{{Abazajian} {et~al.}(2016){Abazajian}, {Adshead}, {Ahmed}, {Allen},
  {Alonso}, {Arnold}, {Baccigalupi}, {Bartlett}, {Battaglia}, {Benson},
  {Bischoff}, {Borrill}, {Buza}, {Calabrese}, {Caldwell}, {Carlstrom}, {Chang},
  {Crawford}, {Cyr-Racine}, {De Bernardis}, {de Haan}, {di Serego Alighieri},
  {Dunkley}, {Dvorkin}, {Errard}, {Fabbian}, {Feeney}, {Ferraro}, {Filippini},
  {Flauger}, {Fuller}, {Gluscevic}, {Green}, {Grin}, {Grohs}, {Henning},
  {Hill}, {Hlozek}, {Holder}, {Holzapfel}, {Hu}, {Huffenberger}, {Keskitalo},
  {Knox}, {Kosowsky}, {Kovac}, {Kovetz}, {Kuo}, {Kusaka}, {Le Jeune}, {Lee},
  {Lilley}, {Loverde}, {Madhavacheril}, {Mantz}, {Marsh}, {McMahon},
  {Meerburg}, {Meyers}, {Miller}, {Munoz}, {Nguyen}, {Niemack}, {Peloso},
  {Peloton}, {Pogosian}, {Pryke}, {Raveri}, {Reichardt}, {Rocha}, {Rotti},
  {Schaan}, {Schmittfull}, {Scott}, {Sehgal}, {Shandera}, {Sherwin}, {Smith},
  {Sorbo}, {Starkman}, {Story}, {van Engelen}, {Vieira}, {Watson}, {Whitehorn},
  \& {Kimmy Wu}}]{cmbs4-sciencebook}
{Abazajian}, K.~N., {Adshead}, P., {Ahmed}, Z., {et~al.}, {CMB-S4 Science Book,
  First Edition}. 2016, arXiv e-prints, \eprint{1610.02743}

\bibitem[{{Ade} {et~al.}(2019){Ade}, {Aguirre}, {Ahmed}, {Aiola}, {Ali},
  {Alonso}, {Alvarez}, {Arnold}, {Ashton}, {Austermann}, \&
  et~al.}]{2019JCAP...02..056A}
{Ade}, P., {Aguirre}, J., {Ahmed}, Z., {et~al.}, {The Simons Observatory:
  science goals and forecasts}. 2019, \jcap, 2, 056, \eprint{1808.07445}

\bibitem[{Aghanim {et~al.}(2020)}]{planck2018likelihood}
Aghanim, N. {et~al.}, {Planck 2018 results. V. CMB power spectra and
  likelihoods}. 2020, Astron. Astrophys., 641, \eprint{1907.12875}

\bibitem[{Aiola {et~al.}(2020)}]{Aiola:2020azj}
Aiola, S. {et~al.}, {The Atacama Cosmology Telescope: DR4 Maps and Cosmological
  Parameters}. 2020, JCAP, 12, 047, \eprint{2007.07288}

\bibitem[{Alonso {et~al.}(2019)Alonso, Sanchez, \& Slosar}]{Alonso2019}
Alonso, D., Sanchez, J., \& Slosar, A., {A unified pseudo-C framework}. 2019,
  Monthly Notices of the Royal Astronomical Society, 484, 4127

\bibitem[{Balkenhol \& Reichardt(2021)}]{balkenhol2021}
Balkenhol, L. \& Reichardt, C.~L. 2021, The Parameter-Level Performance of
  Covariance Matrix Conditioning in Cosmic Microwave Background Data Analyses

\bibitem[{Benoit-L{\'{e}}vy {et~al.}(2013)Benoit-L{\'{e}}vy, D{\'{e}}chelette,
  Benabed, Cardoso, Hanson, \& Prunet}]{Benoit-Levy2013}
Benoit-L{\'{e}}vy, A., D{\'{e}}chelette, T., Benabed, K., {et~al.}, {Full-sky
  CMB lensing reconstruction in presence of sky-cuts}. 2013, \aap,
  \eprint{1301.4145}

\bibitem[{Challinor \& Chon(2004)}]{Challinor2004a}
Challinor, A. \& Chon, G., {Error analysis of quadratic power spectrum
  estimates for CMB polarization: sampling covariance}. 2004, Monthly Notices
  of the Royal Astronomical Society, 360, 509, \eprint{0410097}

\bibitem[{{Chon} {et~al.}(2004){Chon}, {Challinor}, {Prunet}, {Hivon}, \&
  {Szapudi}}]{Chonetal04}
{Chon}, G., {Challinor}, A., {Prunet}, S., {Hivon}, E., \& {Szapudi}, I., {Fast
  estimation of polarization power spectra using correlation functions}. 2004,
  \mnras, 350, 914, \eprint{astro-ph/0303414}

\bibitem[{Dodelson \& Schneider(2013)}]{Dodelson2013}
Dodelson, S. \& Schneider, M.~D., {The effect of covariance estimator error on
  cosmological parameter constraints}. 2013, Physical Review D - Particles,
  Fields, Gravitation and Cosmology, 88, 1, \eprint{1304.2593}

\bibitem[{Dutcher {et~al.}(2021)Dutcher, Balkenhol, Ade, Ahmed, Anderes,
  Anderson, Archipley, Avva, Aylor, Barry, {Basu Thakur}, Benabed, Bender,
  Benson, Bianchini, Bleem, Bouchet, Bryant, Byrum, Carlstrom, Carter, Cecil,
  Chang, Chaubal, Chen, Cho, Chou, Cliche, Crawford, Cukierman, Daley, {De
  Haan}, Denison, Dibert, Ding, Dobbs, Everett, Feng, Ferguson, Foster, Fu,
  Galli, Gambrel, Gardner, Goeckner-Wald, Gualtieri, Guns, Gupta, Guyser,
  Halverson, Harke-Hosemann, Harrington, Henning, Hilton, Hivon, Holder,
  Holzapfel, Hood, Howe, Huang, Irwin, Jeong, Jonas, Jones, Khaire, Knox,
  Kofman, Korman, Kubik, Kuhlmann, Kuo, Lee, Leitch, Lowitz, Lu, Meyer,
  Michalik, Millea, Montgomery, Nadolski, Natoli, Nguyen, Noble, Novosad,
  Omori, Padin, Pan, Paschos, Pearson, Posada, Prabhu, Quan, Raghunathan,
  Rahlin, Reichardt, Riebel, Riedel, Rouble, Ruhl, Sayre, Schiappucci,
  Shirokoff, Smecher, Sobrin, Stark, Stephen, Story, Suzuki, Thompson, Thorne,
  Tucker, Umilta, Vale, Vanderlinde, Vieira, Wang, Whitehorn, Wu, Yefremenko,
  Yoon, \& Young}]{Dutcher2021}
Dutcher, D., Balkenhol, L., Ade, P.~A., {et~al.}, {Measurements of the e -mode
  polarization and temperature- e -mode correlation of the CMB from SPT-3G 2018
  data}. 2021, Physical Review D, 104, 22003, \eprint{2101.01684}

\bibitem[{{Efstathiou}(2004)}]{Efstathiou:2004}
{Efstathiou}, G., {Myths and truths concerning estimation of power spectra: the
  case for a hybrid estimator}. 2004, \mnras, 349, 603,
  \eprint{astro-ph/0307515}

\bibitem[{Friedrich {et~al.}(2021)Friedrich, Andrade-Oliveira, Camacho, Alves,
  Rosenfeld, Sanchez, Fang, Eifler, Krause, Chang, Omori, Amon, Baxter,
  Elvin-Poole, Huterer, Porredon, Prat, Terra, Troja, Alarcon, Bechtol,
  Bernstein, Buchs, Campos, Rosell, Kind, Cawthon, Choi, Cordero, Crocce,
  Davis, DeRose, Diehl, Dodelson, Doux, Drlica-Wagner, Elsner, Everett,
  Fosalba, Gatti, Giannini, Gruen, Gruendl, Harrison, Hartley, Jain, Jarvis,
  MacCrann, McCullough, Muir, Myles, Pandey, Raveri, Roodman, Rodriguez-Monroy,
  Rykoff, Samuroff, S{\'{a}}nchez, Secco, Sevilla-Noarbe, Sheldon, Troxel,
  Weaverdyck, Yanny, Aguena, Avila, Bacon, Bertin, Bhargava, Brooks, Burke,
  Carretero, Costanzi, da~Costa, Pereira, {De Vicente}, Desai, Evrard, Ferrero,
  Frieman, Garc{\'{i}}a-Bellido, Gaztanaga, Gerdes, Giannantonio, Gschwend,
  Gutierrez, Hinton, Hollowood, Honscheid, James, Kuehn, Lahav, Lima, Maia,
  Menanteau, Miquel, Morgan, Palmese, Paz-Chinch{\'{o}}n, Plazas, Sanchez,
  Scarpine, Serrano, Soares-Santos, Smith, Suchyta, Tarle, Thomas, To, Varga,
  Weller, \& Wilkinson}]{Friedrich2021}
Friedrich, O., Andrade-Oliveira, F., Camacho, H., {et~al.}, {Dark energy survey
  year 3 results: Covariance modelling and its impact on parameter estimation
  and quality of fit}. 2021, Monthly Notices of the Royal Astronomical Society,
  3165, 3125, \eprint{2012.08568}

\bibitem[{Garc{\'{i}}a-Garc{\'{i}}a {et~al.}(2019)Garc{\'{i}}a-Garc{\'{i}}a,
  Alonso, \& Bellini}]{Garcia-Garcia2019}
Garc{\'{i}}a-Garc{\'{i}}a, C., Alonso, D., \& Bellini, E., {Disconnected
  pseudo- Cl covariances for projected large-scale structure data}. 2019,
  Journal of Cosmology and Astroparticle Physics, 2019, 043

\bibitem[{Gorski {et~al.}(2005)Gorski, Hivon, Banday, Wandelt, Hansen,
  Reinecke, \& Bartelmann}]{Gorski2005}
Gorski, K.~M., Hivon, E., Banday, A.~J., {et~al.}, {HEALPix: A Framework for
  High-Resolution Discretization and Fast Analysis of Data Distributed on the
  Sphere}. 2005, The Astrophysical Journal, 622, 759, \eprint{0409513}

\bibitem[{Gratton {et~al.}(2022)Gratton, Migliaccio, Challinor, Elsner, Hivon,
  \& Lilley}]{Gratton}
Gratton, S., Migliaccio, M., Challinor, A., {et~al.}, In preparation. 2022

\bibitem[{{Hazumi} {et~al.}(2012){Hazumi}, {Borrill}, {Chinone}, {Dobbs},
  {Fuke}, {Ghribi}, {Hasegawa}, {Hattori}, {Hattori}, {Holzapfel}, {Inoue},
  {Ishidoshiro}, {Ishino}, {Karatsu}, {Katayama}, {Kawano}, {Kibayashi},
  {Kibe}, {Kimura}, {Koga}, {Komatsu}, {Lee}, {Matsuhara}, {Matsumura}, {Mima},
  {Mitsuda}, {Morii}, {Murayama}, {Nagai}, {Nagata}, {Nakamura}, {Natsume},
  {Nishino}, {Noda}, {Noguchi}, {Ohta}, {Otani}, {Richards}, {Sakai}, {Sato},
  {Sato}, {Sekimoto}, {Shimizu}, {Shinozaki}, {Sugita}, {Suzuki}, {Suzuki},
  {Tajima}, {Takada}, {Takagi}, {Takei}, {Tomaru}, {Uzawa}, {Watanabe},
  {Yamasaki}, {Yoshida}, {Yoshida}, \& {Yotsumoto}}]{hazumi2012}
{Hazumi}, M., {Borrill}, J., {Chinone}, Y., {et~al.} 2012, in \procspie, Vol.
  8442, Space Telescopes and Instrumentation 2012: Optical, Infrared, and
  Millimeter Wave, 844219

\bibitem[{{Hivon} {et~al.}(2002){Hivon}, {G{\'o}rski}, {Netterfield}, {Crill},
  {Prunet}, \& {Hansen}}]{Hietal02}
{Hivon}, E., {G{\'o}rski}, K.~M., {Netterfield}, C.~B., {et~al.}, {MASTER of
  the Cosmic Microwave Background Anisotropy Power Spectrum: A Fast Method for
  Statistical Analysis of Large and Complex Cosmic Microwave Background Data
  Sets}. 2002, \apj, 567, 2, \eprint{arXiv:astro-ph/0105302}

\bibitem[{Khersonskii {et~al.}(1988)Khersonskii, Moskalev, \&
  Varshalovich}]{khersonskii1988quantum}
Khersonskii, V., Moskalev, A., \& Varshalovich, D. 1988, Quantum Theory Of
  Angular Momemtum (World Scientific Publishing Company)

\bibitem[{Louis {et~al.}(2020)Louis, Naess, Garrido, \& Challinor}]{Louis2020}
Louis, T., Naess, S., Garrido, X., \& Challinor, A., {Fast computation of
  angular power spectra and covariances of high-resolution cosmic microwave
  background maps using the Toeplitz approximation}. 2020, Physical Review D,
  102, 123538, \eprint{2010.14344}

\bibitem[{Lueker {et~al.}(2010)Lueker, Reichardt, Schaffer, Zahn, Ade, Aird,
  Benson, Bleem, Carlstrom, Chang, Cho, Crawford, Crites, {De Haan}, Dobbs,
  George, Hall, Halverson, Holder, Holzapfel, Hrubes, Joy, Keisler, Knox, Lee,
  Leitch, McMahon, Mehl, Meyer, Mohr, Montroy, Padin, Plagge, Pryke, Ruhl,
  Shaw, Shirokoff, Spieler, Stalder, Staniszewski, Stark, Vanderlinde, Vieira,
  \& Williamson}]{Lueker2010}
Lueker, M., Reichardt, C.~L., Schaffer, K.~K., {et~al.}, {Measurements of
  secondary cosmic microwave background anisotropies with the South Pole
  Telescope}. 2010, Astrophysical Journal, 719, 1045, \eprint{0912.4317}

\bibitem[{Nicola {et~al.}(2021)Nicola, Garc{\'{i}}a-Garc{\'{i}}a, Alonso,
  Dunkley, Ferreira, Slosar, \& Spergel}]{Nicola2021}
Nicola, A., Garc{\'{i}}a-Garc{\'{i}}a, C., Alonso, D., {et~al.}, {Cosmic shear
  power spectra in practice}. 2021, Journal of Cosmology and Astroparticle
  Physics, 2021, \eprint{2010.09717}

\bibitem[{{Planck Collaboration} {et~al.}(2015){Planck Collaboration}, Aghanim,
  Arnaud, Ashdown, Aumont, Baccigalupi, Banday, Barreiro, Bartlett, Bartolo,
  Battaner, Benabed, Beno{\^{i}}t, Benoit-L{\'{e}}vy, Bernard, Bersanelli,
  Bielewicz, Bock, Bonaldi, Bonavera, Bond, Borrill, Bouchet, Boulanger,
  Bucher, Burigana, Butler, Calabrese, Cardoso, Catalano, Challinor, Chiang,
  Christensen, Clements, Colombo, Combet, Coulais, Crill, Curto, Cuttaia,
  Danese, Davies, Davis, de~Bernardis, de~Rosa, de~Zotti, Delabrouille,
  D{\'{e}}sert, {Di Valentino}, Dickinson, Diego, Dolag, Dole, Donzelli,
  Dor{\'{e}}, Douspis, Ducout, Dunkley, Dupac, Efstathiou, Elsner, En{\ss}lin,
  Eriksen, Fergusson, Finelli, Forni, Frailis, Fraisse, Franceschi, Frejsel,
  Galeotta, Galli, Ganga, Gauthier, Gerbino, Giard, Gjerl{\o}w,
  Gonz{\'{a}}lez-Nuevo, G{\'{o}}rski, Gratton, Gregorio, Gruppuso, Gudmundsson,
  Hamann, Hansen, Harrison, Helou, Henrot-Versill{\'{e}},
  Hern{\'{a}}ndez-Monteagudo, Herranz, Hildebrandt, Hivon, Holmes, Hornstrup,
  Huffenberger, Hurier, Jaffe, Jones, Juvela, Keih{\"{a}}nen, Keskitalo,
  Kiiveri, Knoche, Knox, Kunz, Kurki-Suonio, Lagache, L{\"{a}}hteenm{\"{a}}ki,
  Lamarre, Lasenby, Lattanzi, Lawrence, Jeune, Leonardi, Lesgourgues, Levrier,
  Lewis, Liguori, Lilje, Lilley, Linden-V{\o}rnle, Lindholm,
  L{\'{o}}pez-Caniego, Mac{\'{i}}as-P{\'{e}}rez, Maffei, Maggio, Maino,
  Mandolesi, Mangilli, Maris, Martin, Mart{\'{i}}nez-Gonz{\'{a}}lez, Masi,
  Matarrese, Meinhold, Melchiorri, Migliaccio, Millea, Mitra,
  Miville-Desch{\^{e}}nes, Moneti, Montier, Morgante, Mortlock, Mottet, Munshi,
  Murphy, Narimani, Naselsky, Nati, Natoli, Noviello, Novikov, Novikov,
  Oxborrow, Paci, Pagano, Pajot, Paoletti, Partridge, Pasian, Patanchon,
  Pearson, Perdereau, Perotto, Pettorino, Piacentini, Piat, Pierpaoli,
  Pietrobon, Plaszczynski, Pointecouteau, Polenta, Ponthieu, Pratt, Prunet,
  Puget, Rachen, Reinecke, Remazeilles, Renault, Renzi, Ristorcelli, Rocha,
  Rossetti, Roudier, D'Orfeuil, Rubi{\~{n}}o-Mart{\'{i}}n, Rusholme, Salvati,
  Sandri, Santos, Savelainen, Savini, Scott, Serra, Spencer, Spinelli,
  Stolyarov, Stompor, Sunyaev, Sutton, Suur-Uski, Sygnet, Tauber, Terenzi,
  Toffolatti, Tomasi, Tristram, Trombetti, Tucci, Tuovinen, Umana, Valenziano,
  Valiviita, {Van Tent}, Vielva, Villa, Wade, Wandelt, Wehus, Yvon, Zacchei, \&
  Zonca}]{PlanckCollaboration2015}
{Planck Collaboration}, Aghanim, N., Arnaud, M., {et~al.}, {Planck 2015
  results. XI. CMB power spectra, likelihoods, and robustness of parameters}.
  2015, \aap, \eprint{1507.02704}

\bibitem[{{Planck Collaboration} {et~al.}(2020){Planck Collaboration},
  {Aghanim}, {et~al.}}]{Aghanim:2018eyx}
{Planck Collaboration}, {Aghanim}, N., {et~al.}, {Planck 2018 results. VI.
  Cosmological parameters}. 2020, \aap, 641, A6, \eprint{1807.06209}

\bibitem[{Sellentin \& Starck(2019)}]{Sellentin2019}
Sellentin, E. \& Starck, J.~L., {Debiasing inference with approximate
  covariance matrices and other unidentified biases}. 2019, Journal of
  Cosmology and Astroparticle Physics, 2019, \eprint{1902.00709}

\bibitem[{Szapudi {et~al.}(2001)Szapudi, Prunet, Pogosyan, Szalay, \&
  Bond}]{Szapudi2001}
Szapudi, I., Prunet, S., Pogosyan, D., Szalay, A.~S., \& Bond, J.~R., {Fast
  Cosmic Microwave Background Analyses via Correlation Functions}. 2001, The
  Astrophysical Journal, 548, L115

\end{thebibliography}
\bibliographystyle{aat}

%%%%%%%%%%%%%%%%%%%%%%
\newpage
\appendix
%%%
\section{Analysis on the curved sky}
\label{app:Iprop}
This appendix describes the mathematical tools that are used on curved sky for
CMB analysis. We make use of the spherical harmonic decomposition of Gaussian
fields. We introduce various operators that allow us to express the couplings
and the covariance of the power spectra. We make use of some geometrical
relations of spherical harmonics to obtain our results. In this appendix, we
introduce formulae that can be either used in the temperature or the
polarization case.
\label{sec:analysis}

%%%
\subsection{Temperature}
We first consider the case of a map of the CMB intensity anisotropies $T(\hn)$
observed in direction $\hn$. The anisotropies are distributed as a Gaussian
random field with a corresponding power spectrum $C_\ell^\TT$, observed through
a mask $W(\hn)$.

%%%
\subsubsection*{Harmonic coefficients and underlying power spectrum}
The intensity map can be decomposed with spin-0 spherical harmonics to obtain
the harmonic coefficients and their variance, the intensity power spectrum,
which fully characterizes the physical properties of the field
\begin{align}
    a^\T_{\ell m}                                   & = \int \ud \hu
    T(\hu) \ {}_{0}Y^*_{\ell m}(\hu) ,                               \\
    \langle a^\T_{\ell m} a^{\T*}_{\ell'm'} \rangle & =
    C^\TT_{\ell}\delta_{\ell\ell'}
    \delta_{mm'}.
\end{align}
Here the brackets $\langle\rangle$ indicate an average over many realizations of
the maps.

%%%
\subsubsection*{Weighted mask \texorpdfstring{$W(\hn)$}{W}} The weighted mask is
a real map with weights from $0$ to $1$ that is used to taper the data on the
border of the survey area in order to reduce Fourier ringing when using harmonic
decomposition. We define the mask harmonic coefficients and its power spectrum
as
\begin{align}
    w_{\ell m} \equiv \int \ud \hu \label{eq:maskwlm}
    W(\hu) \ {}_{0}Y^*_{\ell m}(\hu) , \\
    \mathcal{W}_\ell \equiv \frac{1}{2\ell+1}\sum_m w_{\ell m}w^*_{\ell m}.
    \label{eq:maskps}
\end{align}

%%%
\subsubsection*{Pseudo-power spectrum estimator}
In \cmb experiments, one way to obtain a biased estimator of the power spectrum
is to define the pseudo-harmonic coefficients and the pseudo-power spectrum
\begin{align}
    \label{eq:psuedoalm_app}
    \ta^\T_{\ell m} & \equiv \int \ud \hu
    W(\hu)T(\hu) \ {}_{0}Y^*_{\ell m}(\hu) , \\
    \label{eq:definepseudoclTT}
    \tC^\TT_\ell    & \equiv
    \frac{1}{2\ell+1} \sum_{m=-\ell}^\ell |\ta^\T_{\ell m}|^2.
\end{align}

%%%
\subsubsection*{Relation between harmonic coefficients}
We relate the masked pseudo-harmonic coefficients to the unmasked one with
\begin{equation}
    \ta^\T_{\ell m} = \sum_{\ell'm'} a^\T_{\ell'm'} \ {}_{0}I_{\ell m\ell'm'}[W].
\end{equation}
The $I[W]$ couplings are defined below and can be expressed in terms of sums
over Wigner-3j symbols and the $w_{\ell m }$, with
\begin{align}
    \label{eq:apppolIkernels}
    {}_{s}I_{\ell m\ell'm'}[W] \equiv \int & \ud\hu {}_{s}Y_{\ell m}(\hu)
    W(\hu) {}_{s}Y^*_{\ell'm'}(\hu),
    \\
    \nonumber
    = \sum_{LM}                            & w_{LM} (-1)^{m'} \left[ \frac{(2\ell+1)(2\ell'+1)(2L+1)}{4\pi} \right]^{1/2} \\
                                           & \begin{pmatrix}
                                                 \ell & \ell' & L \\
                                                 -s   & s     & 0
                                             \end{pmatrix} \times
    \begin{pmatrix}
        \ell & \ell' & L \\
        m    & -m'   & M
    \end{pmatrix}.
    \label{eq:apppolIkernels_wigner}
\end{align}
Here, we anticipated the extension of this notation to the polarized case which
deals with spin-2 fields. The mask-dependent ${}_{s}I_{\ell m\ell'm'}[W]$
coupling coefficients relate the underlying harmonic coefficients to the
measured pseudo-harmonic coefficients. In the full-sky case, one gets
${}_{s}I_{\ell m\ell'm'}[1]=\delta_{\ell \ell'}\delta_{mm'}$ thanks to the
closing relations of the spin-weighted spherical harmonics.

Let us introduce some useful relations that are demonstrated in \cite{Hietal02}
\begin{align}
    \label{eq:relationofI1}
    \sum_{\ell m} {}_{s}I_{\ell_1 m_1\ell m}[W] {}_{s}I^*_{\ell_2m_2\ell m}[W]
                                                                 & = {}_{s}I_{\ell_1 m_1\ell_2m_2}[W^2], \\
    \label{eq:relationofI2}
    \sum_{m_1m_2} \frac{{}_{s}I_{\ell_1 m_1\ell_2m_2}[W]
    {}_{s'}I^*_{\ell_1 m_1\ell_2m_2}[W]}{(2\ell_1+1)(2\ell_2+1)} & =
    \Xi^{ss'}_{\ell_1\ell_2}[W].
\end{align}
Here we introduced the symmetric operator $\Xi^{ss'}$ acting on a power spectrum
$\mathcal{A}_\ell$,
\begin{align}
    \Xi^{ss'}_{\ell\ell'}[\mathcal{A}] \equiv
    \sum_L \frac{2L+1}{4\pi}  \mathcal{A}_L
    \begin{pmatrix}
        \ell & \ell' & L \\
        s    & -s    & 0
    \end{pmatrix}
    \begin{pmatrix}
        \ell & \ell' & L \\
        s'   & -s'   & 0
    \end{pmatrix}.
    \label{eq:definexi}
\end{align}
We extend this definition to an operator acting on a map $A(\hn)$, with
\begin{align}
    \Xi^{ss'}_{\ell\ell'}[A] \equiv \Xi^{ss'}_{\ell\ell'}[\mathcal{A}],
    \label{eq:definexionmap}
\end{align}
where we defined the power spectrum $\mathcal{A}_\ell$ of the map $A$ as in
Eqs.~(\ref{eq:maskwlm}) and (\ref{eq:maskps}).

%%%
\subsubsection*{\master relation between estimated and true spectra}
Inserting Eq.~(\ref{eq:psuedoalm_app}) into Eq.~(\ref{eq:definepseudoclTT}),
using the relations of Eq.~(\ref{eq:relationofI2}) and the definition of
Eq.~(\ref{eq:definexi})-(\ref{eq:definexionmap}), we relate the ensemble average
of the pseudo-power spectrum to the underlying power spectrum using the \master
mode-coupling kernel with
\begin{equation}
    \langle \tC^\TT_\ell\rangle = \sum_{\ell'} {}_0M_{\ell\ell'}
    [W]C^\TT_{\ell'}.
    \label{eq:appendmasterT}
\end{equation}
The \master mode-coupling matrix is given by
\begin{align}
    \label{eq:defineM}
    {}_0 M_{\ell\ell'}[W] & \equiv (2\ell'+1)\Xi^{00}_{\ell \ell'}[W].
\end{align}

%%%
\subsection{Polarization}
We consider the case of the CMB intesity and polarization anisotropies,
represented by maps $T(\hn), Q(\hn), U(\hn)$ in direction $\hn$ of the sky.
These are gaussian random fields, fully characterized by their power spectra
$(C_\ell^\TT, C_\ell^\EE, C_\ell^\BB, C_\ell^\TE)$ observed through a mask
$W(\hn)$.
\label{sec:appmasterpol}
The definitions and relations of the previous section can be extended to
polarization spectra. First we compute the pseudo-harmonic coefficients on the
masked sky with spin weighted spherical harmonics, given in the following
inverse relation from \cite{Chonetal04}
\begin{align}
    (Q\pm i U) (\hn) = \sum_{\ell m }
    (\ta^{\E}_{\ell m} \mp i \ta^{\B}_{\ell m})
    {}_{\mp2}Y_{\ell m }
\end{align}
The pseudo-power spectrum $\tC_\ell^{\rm \textsc{XY}}$ is obtained by summing
over the measured pseudo-harmonic coefficients $\ta^{\rm X}_{\ell m}$, ${\rm
\textsc{X}},{\rm \textsc{Y}} \in [\T, \E, \B]$ with the same multipole $\ell$,
with
\begin{equation}
    \tC^{\rm \textsc{XY}}_\ell =
    \frac{1}{2\ell + 1} \sum_{m=-\ell}^\ell
    \ta^{\rm \textsc{X}}_{\ell m}\ta^{\rm \textsc{Y}*}_{\ell m}
\end{equation}
More details can be found in \cite{Challinor2004a}. Following the same approach
as in the previous section, one can write the \master relation in polarization
\begin{align}
    \langle \tC^\EE_\ell        & + \tC^\BB_\ell \rangle =
    \sum_{\ell'} {}_{+2}M_{\ell\ell'} \left( C^\EE_{\ell'} + C^\BB_{\ell'}\right), \\
    {}_{+2}M_{\ell\ell'}        & = (2\ell'+1) \Xi^{22}_{\ell \ell'}[W],           \\
    \langle \tC^\EE_\ell        & - \tC^\BB_\ell \rangle =
    \sum_{\ell'} {}_{-2}M_{\ell\ell'} \left(C^\EE_{\ell'} - C^\BB_{\ell'}\right),  \\
    {}_{-2}M_{\ell\ell'}        & = (2\ell'+1) \Xi^{2-2}_{\ell \ell'}[W],          \\
    \langle \tC^\TE_\ell\rangle & =
    \sum_{\ell'} {}_{\times}M_{\ell\ell'} C^\TE_{\ell'},                           \\
    {}_{\times}M_{\ell\ell'}    & = (2\ell'+1) \Xi^{20}_{\ell \ell'}[W].
\end{align}

%%%
\subsection{Renormalised kernels}
\label{sec:apprenormalised}
We define the renormalised \master kernels, which will be used in the \INKA
approximation. They are written as
\begin{align}
    {}_{\texttt{k}}\bar{M}_{\ell\ell'} \equiv \frac{1}{\sum_{\ell'}
    {}_{\texttt{k}}M_{\ell\ell'}} {}_{\texttt{k}}M_{\ell\ell'}, \ \ \forall
    \texttt{k} \in [0, -2, +2, \times] \label{eq:renormalised}.
\end{align}
Summing over $\ell'$ yields
\begin{equation}
    \sum_{\ell'} {}_{\texttt{k}}\bar{M}_{\ell\ell'} = 1,
\end{equation}
which ensures that the approximated covariance coupling kernel defined in
Eq.~(\ref{eq:definka}) is properly normalized.

%%%
\section{Covariance of the pseudo-power spectrum}
\label{app:covar}
In this appendix, we outline how the formula of the covariance matrix of the
pseudo-power spectrum is obtained, in the temperature case. Our goal is to
introduce Eq.~(\ref{eq:finalappB}). The covariance matrix of the pseudo-power
spectrum writes
\begin{align}
    \tilde{\Sigma}_{\ell \ell'} & = \langle \tC_\ell \tC_{\ell '} \rangle -
    \langle \tC_\ell \rangle \langle \tC_{\ell '} \rangle,              \nonumber \\
                                & =  \sum_{mm'}\frac{
        \langle |\ta_{\ell m}|^2
        |\ta_{\ell ' m'}|^2 \rangle
        - \langle |\ta_{\ell m}|^2 \rangle
        \langle |\ta_{\ell ' m'}|^2 \rangle
    }{(2\ell+1)(2\ell'+1)}.
\end{align}
As the intensity map $T(\hn)$ is real and the spherical harmonics follow
\begin{align}
    {}_{0}Y_{\ell m}^* = (-1)^m {}_{0}Y_{\ell(-m)},
\end{align}
the spherical harmonic coefficients of $T(\hn)$ follow
\begin{align}
    a_{\ell m}^* = (-1)^m a_{\ell(-m)}.
\end{align}

Then, when computing the four-point function, and thanks to Wick's theorem,
\begin{align}
    \sum_{mm'} & \langle |\ta_{\ell m}|^2
    |\ta_{\ell'm'}|^2 \rangle - \langle |\ta_{\ell m}|^2 \rangle
    \langle |\ta_{\ell'm'}|^2 \rangle            \nonumber             \\
    =          & \sum_{mm'} \langle \ta_{\ell m} \ta_{\ell'm'} \rangle
    \langle \ta^*_{\ell m}\ta^*_{\ell'm'} \rangle +
    \langle \ta_{\ell m} \ta^*_{\ell'm'} \rangle
    \langle \ta^*_{\ell m}\ta_{\ell'm'} \rangle.
\end{align}
Looking at the first term of this sum, and using the change of variable
$m''=-m'$, it is straightforward to show that
\begin{align}
    \sum_{mm'} & \langle \ta_{\ell m} \ta_{\ell'm'} \rangle
    \langle \ta^*_{\ell m}\ta^*_{\ell'm'} \rangle               \nonumber                 \\
               & = \sum_{mm'} (-1)^{(2m')}\langle \ta_{\ell m} \ta^*_{\ell'(-m')} \rangle
    \langle \ta^*_{\ell m}\ta_{\ell'(-m')} \rangle,         \nonumber                     \\
               & = \sum_{mm''} \langle \ta_{\ell m} \ta^*_{\ell'm''} \rangle
    \langle \ta^*_{\ell m}\ta_{\ell'm''} \rangle.
\end{align}
Finally, one has
\begin{align}
    \sum_{mm'} & \langle |\ta_{\ell m}|^2\nonumber
    |\ta_{\ell'm'}|^2 \rangle - \langle |\ta_{\ell m}|^2 \rangle
    \langle |\ta_{\ell'm'}|^2 \rangle,                                            \\
               & = 2 \sum_{mm'} \langle \ta_{\ell m} \ta^*_{\ell'm'} \rangle
    \langle \ta^*_{\ell m}\ta_{\ell'm'} \rangle    \nonumber                      \\
               & = 2 \sum_{mm'} |\langle \ta_{\ell m} \ta^*_{\ell'm'} \rangle|^2.
\end{align}
Then, one has, for the covariance matrix,
\begin{align}
    \tilde{\Sigma}_{\ell \ell'} & = \frac{2}{(2\ell+1)(2\ell'+1)} \sum_{mm'}
    |\langle \ta_{\ell
        m} \ta^*_{\ell ' m'} \rangle|^2.
    \label{eq:appdefpseudocov}
\end{align}
Using the decomposition of pseudo-harmonic coefficients, yields
\begin{align}
    \langle \ta_{\ell m} \ta^*_{\ell ' m'} \rangle
     & = \sum_{\ell_1m_1\ell_2m_2} \langle a_{\ell_1m_1}a^*_{\ell_2m_2} \rangle
    I_{\ell m\ell_1m_1}[W]I^*_{\ell'm'\ell_2m_2}[W],     \nonumber              \\
     & = \sum_{\ell_1m_1} C_{\ell_1}
    I_{\ell m\ell_1m_1}[W]I^*_{\ell'm'\ell_1m_1}[W].
    \label{eq:2pt}
\end{align}
Inserting Eq.~(\ref{eq:2pt}) into Eq.~(\ref{eq:appdefpseudocov}) gives
\begin{align}
    \label{eq:finalappB}
    \tilde{\Sigma}_{\ell \ell'} = & \frac{2}{(2\ell+1)(2\ell'+1)} \sum_{mm'}
    \sum_{\ell_1m_1} \sum_{\ell_2m_2}
    C_{\ell_1}C_{\ell_2}                                                            \\ \nonumber
                                  & I_{\ell m\ell_1m_1}[W]I^*_{\ell'm'\ell_1m_1}[W]
    I^*_{\ell m\ell_2m_2}[W]I_{\ell'm'\ell_2m_2}[W].
\end{align}

%%%
\section{Expansion in Legendre series}
\label{sec:applegendre}
In this section, we introduce the Legendre transforms of the harmonic quantities
used in this work. Each relation in harmonic space has a corresponding
expression in real space. The \polspice software relies on the relations in real
space.

%%%
\subsection{From spin-0 spectra to two point correlation functions}
Given a spectrum $\mathcal{A}_\ell$, one can associate with it a real two-point
correlation function $a$
\begin{equation}
    \label{eq:appspectocorr}
    a(\theta) =
    \sum_\ell \frac{2\ell+1}{4\pi} \mathcal{A}_\ell P_\ell(\cos\theta)
    \ \ \forall \theta \in [0, \pi].
\end{equation}
The inverse relation is
\begin{equation}
    \mathcal{A}_\ell = 2\pi \int_0^\pi \ud\theta \sin
    \theta a(\theta) P_\ell(\cos \theta)
    \ \ \forall \ell \geq 0.
\end{equation}
The two-point function gives the correlations between two directions of the sky,
for instance in the \cmb anisotropies full sky case, we can write
\begin{align}
    \langle T(\hn_1)T(\hn_2)\rangle
    = \sum_\ell \frac{2\ell+1}{4\pi}
    C^\mathrm\TT_\ell P_\ell(\hn_1\cdot \hn_2).
\end{align}
\subsection{From convolution to multiplication}
A convolution with a square matrix $A$ in harmonic space, such as in
Eq.~(\ref{eq:appendmasterT}) is equivalent to a multiplication in real space
with the correlation function $a(\theta)$ given by
\begin{align}
    \label{eq:mattocorr}
    a(\theta)\frac{2\ell'+1}{4\pi}P_{\ell'}(\cos\theta)
    = \sum_{\ell} \frac{2\ell+1}{4\pi}A_{\ell\ell'} \ \
    \forall \theta \in [0, \pi].
\end{align}
The inverse relation is
\begin{equation}
    A_{\ell\ell'} \equiv \frac{2\ell'+1}{2} \int_0^{\pi}
    a(\theta) P_{\ell}(\cos \theta) P_{\ell'}(\cos \theta) \sin(\theta)\ud \theta.
\end{equation}
The last relation is equivalent to
\begin{equation}
    A_{\ell\ell'} = (2\ell'+1) \Xi^{00}_{\ell\ell'}[\mathcal{A}],
\end{equation}
with $\mathcal{A}$ the power spectrum associated with the two-point function $a$
through a Legendre transform. We can thus extend the definition of the operator
$\Xi$ to an operator acting on a correlation function $a$, with
\begin{align}
    \Xi^{ss'}_{\ell\ell'}[a] \equiv \Xi^{ss'}_{\ell\ell'}[\mathcal{A}].
    \label{eq:definexioncorrfunc}
\end{align}
Here we have already extended the definition to be used in the spin-2 case,
which we discuss in the next subsection.

%%%
\subsection{Spin-2}
Similar rules to the ones introduced above can be written for spin-2 quantities,
by replacing the Legendre polynomials by more general reduced Wigner $d$-matrix
$d^\ell_{2\pm2}$. Details can be found in \cite{Challinor2004a}. The spin-2
relations are, for the $\mathcal{A}^{\pm}_\ell$ power spectra associated with a
spin-2 field
\begin{align}
    a_{\pm}(\theta)        & =
    \sum_\ell \frac{2\ell+1}{4\pi} \beta_\ell d^\ell_{2\pm2}(\cos\theta), \\
    \label{eq:corrtospecspin2}
    \mathcal{A}^{\pm}_\ell & = 2\pi \int_0^\pi \ud\theta \sin
    \theta a_{\pm}(\theta) d^\ell_{2\pm2}(\cos \theta).
\end{align}
We can associate a correlation function with its spin-2 convolution matrix,
\begin{align}
    {}_{\pm2}A_{\ell\ell'} & = \frac{2\ell'+1}{2} \int_0^{\pi}
    a(\theta) d^\ell_{2\pm2}(\cos \theta) d^{\ell'}_{2\pm2}(\cos \theta)
    \sin(\theta)\ud \theta,             \nonumber                     \\
                           & = (2\ell'+1) \Xi^{2\pm2}_{\ell\ell'}[a].
\end{align}
We can also compute the matrix associate with the spin-0 cross spin-2 case
\begin{align}
    {}_{\times}A_{\ell\ell'}
     & = (2\ell'+1) \Xi^{20}_{\ell\ell'}[a].
\end{align}

%%%
\subsection{Applying this formalism to the \master matrix}
In our case, we can write, for $s \in [0, 2]$,
\begin{align}
    {}_{\pm s}M_{\ell\ell'} = (2\ell'+1) \Xi^{s\pm s}_{\ell\ell'}[w]
    \label{eq:mastermatrixtocorr}
\end{align}
with $w(\theta)$ the correlation function of the mask.

Let us apply the previous formalism and particularly the
Eq.~(\ref{eq:mastermatrixtocorr}) to the \master relation. We use Legendre
series expansion of the true power spectrum $C_\ell$ and the pseudo-power
spectrum estimator $\tC_\ell$ to define the correlation functions $\xi$ and
$\tilde{\xi}$ respectively. It gives
\begin{align}
    \tilde{\xi}(\theta)
     & \equiv \sum_\ell \frac{2\ell+1}{4\pi}P_\ell(\cos \theta)\tC_\ell, \\
    \xi(\theta)
     & \equiv \sum_\ell \frac{2\ell+1}{4\pi}P_\ell(\cos \theta)C_\ell.
\end{align}
Starting from the right hand side of Eq.~(\ref{eq:masterT}) and going to real
space using a Legendre transform at an angle $\theta \in [0, \pi]$, we have
\begin{align}
    \sum_{\ell\ell'} \frac{2\ell+1}{4\pi} & P_\ell(\cos\theta)
    {}_0M_{\ell\ell'} C_{\ell'} \nonumber                                                  \\
                                          & = \sum_{\ell'} w(\theta) \frac{2\ell'+1}{4\pi}
    P_{\ell'}(\cos\theta)C_{\ell'},           \nonumber                                    \\
                                          & = w(\theta) \xi(\theta),   \nonumber           \\
    \text{which implies } \langle \tilde{\xi}(\theta) \rangle
                                          & = w(\theta) \xi(\theta).
\end{align}
with $w(\theta)$ the correlation function of the mask.

\subsection{\polspice in polarization}
This section aims at describing the regularization technique used for
polarization by \polspice. One of the main advantages of \polspice is that it
allows the possibility of eliminating \EE \ to \BB  \ (and \BB \ to \EE) mixing,
using non-local relations between Wigner $d$-matrices, see Sec.~5 of
\cite{Chonetal04}. The obtained estimator $\hC_{\ell}^\EE$ (respectively
$\hC_{\ell}^\BB$) depends only on the average of $C_{\ell}^\EE$ (respectively
$C_{\ell}^\BB$)  and the scalar apodizing function $f$.

Using the Legendre transforms of spin-2 quantities
(Eq.~(\ref{eq:corrtospecspin2})), let us associate the correlation functions
$\xi_\pm$ with the spectra $C_\ell^\EE \pm C_\ell^\BB$ and $\tilde{\xi}_{\pm}$
to $\tC_\ell^\EE \pm \tC_\ell^\BB$. \polspice builds two correlation functions
$\hat{\xi}_{\mathrm{dec}}$ and $\hat{\xi}_{-}$ to produce an estimator of the
true underlying polarized power spectrum. The latter is defined similarly to
$\hat{\xi}$ in Eq.~(\ref{eq:polspicecorr}),
\begin{align}
    \label{eq:minusspicecorrtominustildecorr}
    \hat{\xi}_{-}(\theta) = g(\theta)\tilde{\xi}_{-}(\theta).
\end{align}
The first is built on integral relations. \polspice eliminates the mixing
inherent in $\tilde{\xi}_{+}$ with the following relation in real space,
\begin{align}
    \label{eq:decspicecorrtoplustildecorr}
    \hat{\xi}_{\mathrm{dec}}(\theta) = g(\theta) \int_{-1}^1 \ud \cos \theta'
    \tilde{\xi}_{+}(\theta')  \sum_\ell d^\ell_{2-2}(\theta)d^\ell_{22}(\theta').
\end{align}
This correlation function is noted with the subscript \emph{dec} to emphasize
that it is the crucial step allowing the decoupling of the polarized estimator.
When averaging out Eq.~(\ref{eq:minusspicecorrtominustildecorr}) and
Eq.~(\ref{eq:decspicecorrtoplustildecorr}) on multiple realizations,
\begin{align}
    \langle \hat{\xi}_{\mathrm{dec}} (\theta) \rangle
     & = f(\theta) \int_{-1}^1 \ud \cos \theta'
    \xi_{+}(\theta')  \sum_\ell d^\ell_{2-2}(\theta)d^\ell_{22}(\theta'),
    \label{eq:deccorrtotrue}                    \\
    \langle \hat{\xi}_{-}  (\theta)\rangle
     & = f(\theta) \xi_{-}(\theta).
    \label{eq:minuscorrtotrue}
\end{align}
In harmonic space, transforming Eq.~(\ref{eq:deccorrtotrue}) and
Eq.~(\ref{eq:minuscorrtotrue}) using $d^\ell_{2-2}$, it gives, with
${}_{-2}K_{\ell\ell'} = (2\ell'+1)\Xi^{2-2}_{\ell\ell'}[f^\mathrm{apo}]$,
\begin{align}
    \hC_\ell^\EE + \hC_\ell^\BB                   & \equiv
    2\pi \int_0^\pi \ud\cos \theta \
    \hat{\xi}_{\mathrm{dec}} (\theta) d^\ell_{2-2}(\theta),     \\
    \hC_\ell^\EE - \hC_\ell^\BB                   & \equiv
    2\pi \int_0^\pi \ud\cos \theta \
    \hat{\xi}_{-} (\theta) d^\ell_{2-2}(\theta),                \\
    \text{which implies}                          & \ \nonumber \\
    \langle \hC_\ell^\EE \pm \hC_\ell^\BB \rangle & =
    \sum_{\ell '}{}_{-2}K_{\ell\ell'} (C_{\ell'}^\EE \pm C_{\ell'}^\BB).
\end{align}
Summing or subtracting the last equation for $+$ or $-$ allows one to build
unmixed estimators of the polarization power spectra. If one had chosen not to
decouple the correlation functions and had built $\hC_\ell'^\EE + \hC_\ell'^\BB$
as the Legendre transform of $\hat{\xi}_+ = g\tilde{\xi}_+$, the output
\polspice spectra would follow
\begin{align}
    \langle \hC_\ell'^\EE \pm \hC_\ell'^\BB \rangle =
    \sum_{\ell '}
    {}_{\pm2}K_{\ell\ell'} (C_{\ell'}^\EE \pm C_{\ell'}^\BB),
\end{align}
which would leave some mixing in the polarization spectra, given that ${}_{+2}K
    \neq {}_{-2}K$.

%%%
\section{Multi-mask analysis}
In this section, we will generalize our analysis to multiple masks. This
situation occurs when performing cross-power spectrum analysis with maps with
different masks. We shall restrict ourselves to the study of the intensity case.
Writing the expression of the covariance explicitly as in
\cite{Garcia-Garcia2019}, one has the following expression
\begin{alignat*}{3}
    \  & \cov(\tC_\ell^{i, j}, \tC_{\ell'}^{p, q})             &  &         \\
    \  & = \frac{1}{(2\ell+1)(2\ell'+1)}    \sum_{mm'}         &  & \left [
    \langle \ta^i_{\ell m } \ta^{p*}_{\ell'm' } \rangle \langle \ta^{j*}_{\ell m
    } \ta^{q}_{\ell'm' } \rangle + (p\leftrightarrow q) \right],            \\
    \  & = \frac{1}{(2\ell+1)(2\ell'+1)}   \sum_{\ell_1\ell_2} &  &
    \sum_{m_1m_2mm'} \left [ {C}_{\ell_1}^{i,p} I_{k k_1}[W^i]I_{k_1k'}[W^p]
    \right.                                                                 \\
       & \                                                     &  & \left.
    {C}_{\ell_2}^{j,q} I_{k_2k}[W^j]I_{k'k_2}[W^q] + (p\leftrightarrow q) \right
    ],                                                                      \\
    \  & = \Xi_{\ell\ell'}^{00}[W^i, W^p,W^j, W^q]             &  &
    \sum_{\ell_1\ell_2} \left[ {C}^{i,p}_{\ell_1}
    \bar{\Theta}_{\ell\ell'}^{\ell_1\ell_2}[W^i, W^p,W^j, W^q]
    {C}^{j,q}_{\ell_2} \right. +                                            \\
    \  & \                                                     &  &
        \left.(p\leftrightarrow q) \right].  \label{eq:multiplemask}
\end{alignat*}
We noted $k_i=(\ell_i,m_i)$. We also extended the definition of the kernels to
the case multiple masks as
\begin{align}
     & \Xi_{\ell\ell'}^{00}[W^i,W^p)(W^j,W^q]
    \equiv \Xi_{\ell\ell'}^{00}[\mathcal{V}^{(ip)\times(jq)}], \\
     & \text{where } \mathcal{V}^{(ip)\times(jq)}_\ell \equiv
    \frac{1}{2\ell+1}\sum_{m}\left[W^iW^p\right]_{\ell m }
    \left[W^jW^q\right]^*_{\ell m },                           \\
    \label{eq:thetamultiplemasks}
     & \Theta_{\ell\ell'}^{\ell_1\ell_2} [W^i, W^p,W^j, W^q]=
    \sum_{mm'm_1m_2}
    I_{\ell m\ell_1m_1}[W^i]                                   \\
     & \ I_{\ell_1m_1\ell'm'}[W^p] \
    \nonumber
    I_{\ell'm'\ell_2m_2}[W^j] \ I_{\ell_2m_2\ell m}[W^q].
\end{align}
As long as the masks considered have similar properties, for instance that none
of them include point sources, the computations developed in this work will
hold. Indeed, in Eq.~(\ref{eq:Xexpan}), the assumptions hold as long as the mask
harmonic coefficients fall quickly enough, which is the case even if the survey
area varies a little from one map to another.

%%%
\section{Details on the \ACC approximation}
\label{sec:appacc}
\subsection{Mathematical validation}
Let us explore why the mathematical justification of the \ACC approximation.
From Fig.~\ref{fig:allfourc}, the  $\bThe_{\ell\ell'}^{\ell_1\ell_2}$ kernel
seems to only depend on $\Delta\equiv\ell'-\ell$, $\Delta_1\equiv\ell_1-\ell$
and $\Delta_2\equiv\ell_2-\ell$. We recall that the normalization of the reduced
coupling kernel (Eq.~(\ref{eq:norm_fourcX})) already approximately only depends
on $\Delta$, since $\Xi^{00}_{\ell\ell'}$ is close to a Toeplitz matrix
\citep{Louis2020}. The kernel itself is given by a summation of products of four
coupling coefficients ${}_{0}I_{\ell m\ell_1m_1}$. Those are expressed as the
sum of the mask window function with a product of two Wigner-3j symbols, as
shown in Eq.~(\ref{eq:apppolIkernels_wigner}). One can remark that since the
mask power spectrum falls relatively fast (Fig.~\ref{fig:maskonsphere}), mostly
the terms with low $L$ in the sum in Eq.~(\ref{eq:apppolIkernels_wigner}) will
contribute. However, we are interested in the cases where all of the other
multipoles $\ell, \ell', \ell_1$ and $\ell_2$ are significantly larger than the
width of the mask spectrum. For this reason, all of the Wigner-3j symbols in
Eq.~(\ref{eq:apppolIkernels_wigner}) can be replaced by their asymptotic
behavior, where in the limit $\ell_i,\ell_j \gg L_i$, we have
\begin{equation}
    \label{eq:approxwigner}
    \begin{pmatrix}
        \ell_i & \ell_j & L_i \\
        m_i    & m_j    & M_i
    \end{pmatrix}
    \approx
    \frac{(-1)^{\ell_j+m_j}}{\sqrt{2\ell_j+1}}d^{L_i}_{M_i, (\ell_j-\ell_i)}
    (\theta),
\end{equation}
\citep{khersonskii1988quantum}. Here, $\theta =\arccos(
    -m_j/(\ell_j(\ell_j+1))^{1/2})$ and $d^{j}_{k, m}$ are reduced Wigner
    rotation matrices.

Introducing this approximation in Eq.~(\ref{eq:apppolIkernels_wigner}),
Eq.~(\ref{eq:deffourcX}) now reads
\begin{align}
    \Theta_{\ell_3\ell_4}^{\ell_1\ell_2}
    \label{eq:Xexpan}
     & \approx \frac{1}{{(4\pi)^2}}
    \sum_{L_iM_i} \Pi_i  w_{L_iM_i} \sqrt{2L_i+1}
    d^{L_i}_{0, (\ell_{i+1}-\ell_{i})}(\pi/2)                \\ \nonumber
     & \left[\sum_{m_i} d^{L_i}_{M_i, (\ell_{i+1}-\ell_{i})}
    \left(\arccos\frac{-m_{i+1}}{(\ell_{i+1}(\ell_{i+1}+1))^{1/2}}\right) \right].
\end{align}
Here we have defined $\ell_{5}\equiv\ell_1$ for notational purposes. We note
that, when $\ell_i$ are large enough, which is the case since $\ell_i,\ell_j \gg
L_i$, $\arccos\frac{-m_{i+1}}{(\ell_{i+1}(\ell_{i+1}+1))^{1/2}}$ explore the
$[0,\pi]$ range, and the expression in brackets in the last equation can be seen
as a Riemann sum over $\theta \in [0,\pi]$. This expression can be approximated
by the integral $\int d^{L_i}_{M_i, (\ell_{i+1}-\ell_{i})}(\theta) \ud\theta$
which only depends on $L_i, M_i$ and $\ell_{i+1}-\ell_{i}$. Since the $L_i, M_i$
are summed over in Eq.~(\ref{eq:Xexpan}), we directly see that as soon as $\ell,
\ell', \ell_1$ and $\ell_2$ are significantly larger than the width of the mask
spectrum, the coupling kernel behaves as a function of their difference, and we
expect that this approximation improves in accuracy with $\ell$.

\subsection{\EE expression}
For now, if we ignore \BB to \EE leakage, we can write
\begin{align}
    \tSig^\EEEE_{\ell\ell'} & \approx 2
    \Xi^{22}_{\ell\ell'}[W^2]
    \left[ C^\EE \cdot \bThe_{\ell\ell'}^{++++}[W]  \cdot  C^\EE \right]
    \label{eq:accee}
\end{align}
where we defined the polarized covariance coupling
\begin{align}
    \Theta_{\ell\ell'}^{\ell_1\ell_2;++++} =
    \sum_{mm_1m'm_2}
    {}_{+}I_{\ell m\ell_1m_1} \ {}_{+}I_{\ell_1m_1\ell'm'} \
    {}_{+}I_{\ell'm'\ell_2m_2} \ {}_{+}I_{\ell_2m_2\ell m}.
\end{align}
Thanks to relations in \citet{Challinor2004a}, this kernel $\Theta^{++++}$ can
be approximately normalized to
\begin{align}
    \sum_{\ell_1\ell_2} \Theta_{\ell\ell'}^{\ell_1\ell_2;++++}[W] \approx
    \Xi_{\ell\ell'}^{22}[W^2],
\end{align}
hence the relation (\ref{eq:accee}).

%%%%%%%%%%%%%%%%%%%%%%

\end{document}